\documentclass[9pt,conference]{IEEEtran}
\IEEEoverridecommandlockouts
\usepackage{cite}
\usepackage{amsmath,amsthm,amssymb,amsfonts}
\usepackage{algorithmic}
\usepackage{graphicx}
\usepackage{textcomp}
\usepackage{xcolor}
\usepackage[caption=false,font=footnotesize,labelfont=rm,textfont=rm]{subfig}
\usepackage{siunitx}
\usepackage{makecell,multirow}
\usepackage{booktabs}
\usepackage{geometry}
\usepackage{cite}
\IEEEaftertitletext{\vspace{-1.8\baselineskip}}

\def\BibTeX{{\rm B\kern-.05em{\sc i\kern-.025em b}\kern-.08em
    T\kern-.1667em\lower.7ex\hbox{E}\kern-.125emX}}

\begin{document}

\title{\vspace{-0.3cm}\huge
ATSim3D: Towards Accurate Thermal Simulator for Heterogeneous 3D-IC Systems Considering Nonlinear Leakage and Conductivity}
\vspace{-0.5cm}

\author{\IEEEauthorblockN{Qipan Wang}
\IEEEauthorblockA{
    \textit{School of Integrated Circuits, Peking University} \\ \textit{Academy for Advanced Interdisciplinary Studies} \\
    qpwang@pku.edu.cn
}
\and
\IEEEauthorblockN{Tianxiang Zhu}
\IEEEauthorblockA{
    \textit{School of Integrated Circuits} \\ \textit{Peking University} \\
    txzhu@pku.edu.cn
}
\and
\IEEEauthorblockN{Yibo Lin\footnotemark*}
\IEEEauthorblockA{
    \textit{School of Integrated Circuits, Peking University} \\ 
    \textit{Institute of Electronic Design Automation, Wuxi} \\
    yibolin@pku.edu.cn
}
\and
\IEEEauthorblockN{Runsheng Wang}
\IEEEauthorblockA{
    \textit{School of Integrated Circuits, Peking University} \\ 
    \textit{Institute of Electronic Design Automation, Wuxi} \\
    r.wang@pku.edu.cn
}
\and
\IEEEauthorblockN{Ru Huang}
\IEEEauthorblockA{
    \textit{School of Integrated Circuits, Peking University} \\ 
    \textit{Institute of Electronic Design Automation, Wuxi} \\
    ruhuang@pku.edu.cn
}
}
\maketitle

\renewcommand{\thefootnote}{\fnsymbol{footnote}}
\footnotetext[1]{Corresponding Author}

\begin{abstract}
    Thermal simulation plays a fundamental role in the thermal design of integrated circuits, especially 3D-ICs. Current simulators require significant runtime for high-resolution simulation, and dismiss the complex nonlinear thermal effects, such as nonlinear thermal conductivity and leakage power. To address these issues, we propose ATSim3D, a thermal simulator for simulating the steady-state temperature profile of nonlinear and heterogeneous 3D-IC systems. We utilize the global-local approach, combining a compact thermal model at the global level, and a finite volume method at the local level. We tackle the nonlinear effects with Kirchhoff transformation and iteration. ATSim3D enables local-level parallelization that helps achieve an average speedup of 40× compared to COMSOL, with a relative error $<3\%$ and a state-of-the-art resolution of 4096 × 4096, holding promise for enhancing thermal-aware design in 3D-ICs.

\end{abstract}


\section{Introduction}
Ever-increasing integration and power consumption of integrated circuits (ICs) lead to severe thermal issues, deteriorating the performance and reliability of modern chips \cite{pedram2006thermal}. Thus, thermal issues have become one of the most crucial factors in VLSI chip design and runtime management. Thermal simulation serves as an essential element to enable thermal optimization and management.

Nowadays, many thermal simulation algorithms have been proposed \cite{sultan2019survey}. 
Among them, numerical methods, encompassing finite difference methods (FDM), finite volume methods (FVM), and finite element methods (FEM), receive popularity for their high accuracy and applicability to various scenarios. Hence commercial thermal simulators are usually based on discretization-based numerical methods, especially FEM, including COMSOL \cite{comsol}, Celsius \cite{Celcius}, etc.
However, FEM is computationally expensive and memory intensive. To overcome the efficiency issue, many open-sourced simulators like HotSpot \cite{stan2003hotspot} and 3D-ICE \cite{terraneo20213d}, construct and solve compact thermal models (RC network) based on FDM. One step further, PACT \cite{yuan2021pact} exploits a SPICE solver for parallel high-resolution simulation.

However, we identify there exist three challenges in existing simulators. 
Firstly, an efficient fine-grained (standard-cell level) simulator is still missing. Most simulators feature grid resolutions ranging from 32$\times$32 to 256$\times$256 (block level), while thermal-aware physical design requires partitioning the active layers into 1024$\times$1024 grids and beyond \cite{yuan2021pact,Nanoheat}, since modern ICs have reached sizes of several square millimeters. Although PACT can simulate with the state-of-the-art (SOTA) resolution of 1024$\times$1024, it has already reached the limit and can hardly further scale up with grid sizes.

\begin{figure}[tb]
\vspace{-0.4cm}
\centering
\subfloat[Monolithic 3D-IC]{
\begin{minipage}[t]{0.46\linewidth}
    \centering
    \includegraphics[width=1\linewidth]{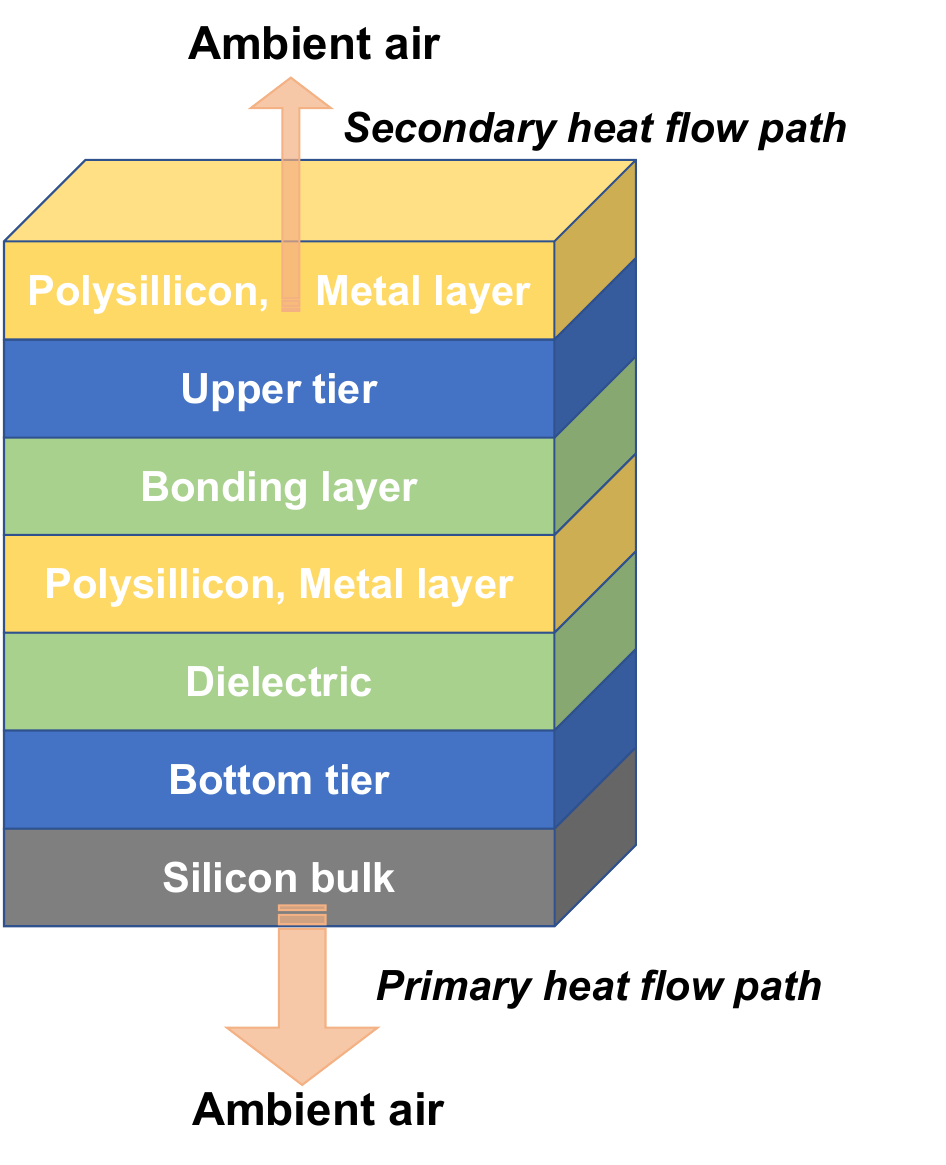}
\end{minipage}
}
\subfloat[TSV-based 3D-IC]{
\begin{minipage}[t]{0.56\linewidth}
    \centering
    \includegraphics[width=1\linewidth]{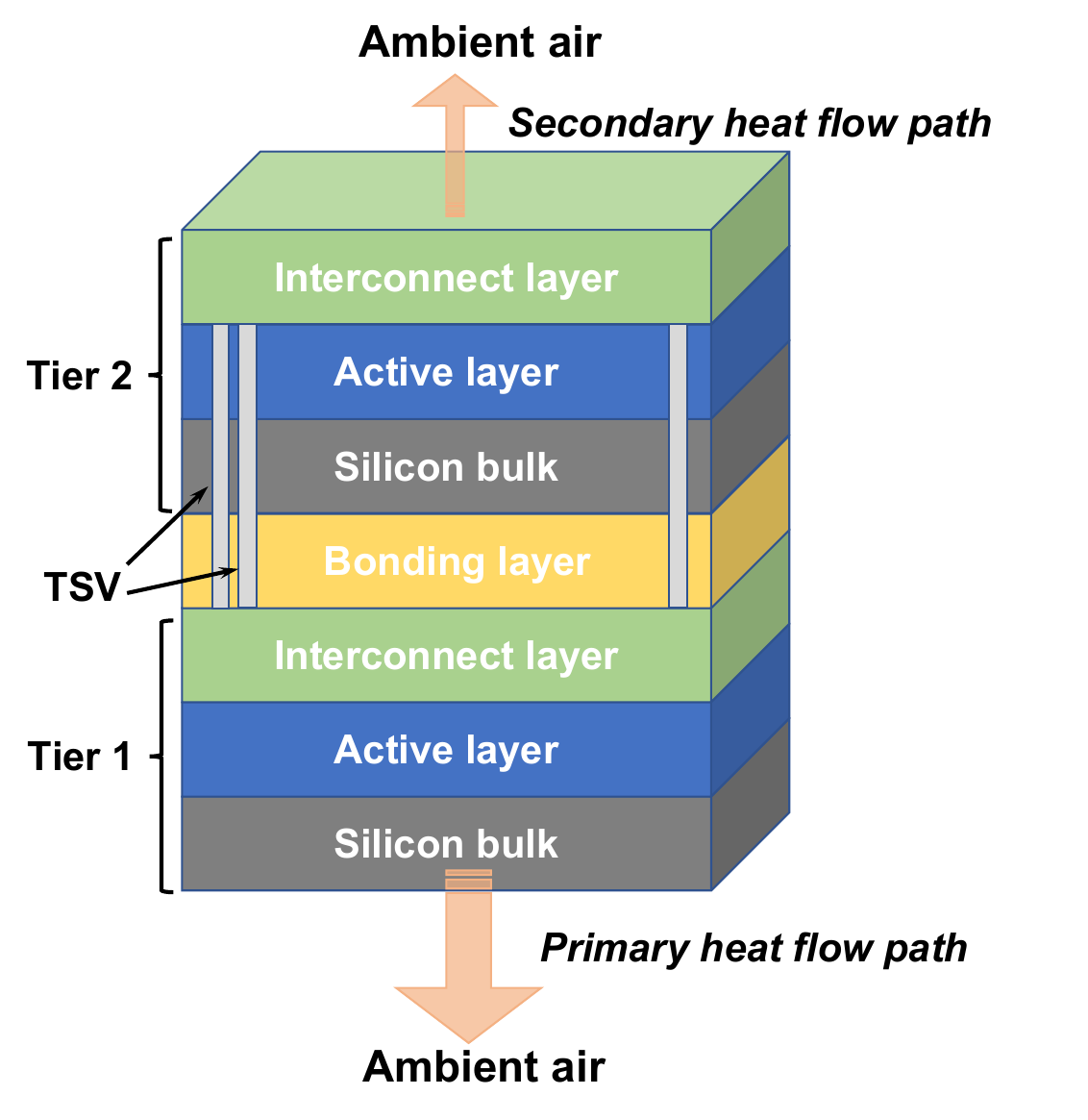}
\end{minipage}
}
\centering
\vspace{-0.15cm}
\caption{Cross-section of two 3D-IC structures for (a) a Mono3D-IC with 2 active layers, (b) a two-tier TSV-based 3D-IC.}\label{Fig:3DIC}
\vspace{-0.3cm}
\end{figure}

The second challenge lies in the heterogeneous nature of real-world chips, especially 3D-ICs. As shown in Fig. \ref{Fig:3DIC}, 3D-ICs stack multiple active layers and can provide better performance than 2D counterparts \cite{shukla2019overview}. Through silicon vias (TSV), composed of copper or tungsten, is introduced in 3D-ICs for conducting signal and power, as shown in Fig. \ref{Fig:3DIC}(b). 
Previous chip-level thermal simulators often assume uniform materials within each layer and struggle to simulate the heterogeneous systems, especially when the TSV radius is small ($5\sim20\mu m$ typically). 

\begin{figure}[tb]
\vspace{-0.5cm}
\centering
\subfloat[nonlinear conductivity of Si]{
\begin{minipage}[t]{0.48\linewidth}
    \centering
    \includegraphics[width=0.95\linewidth]{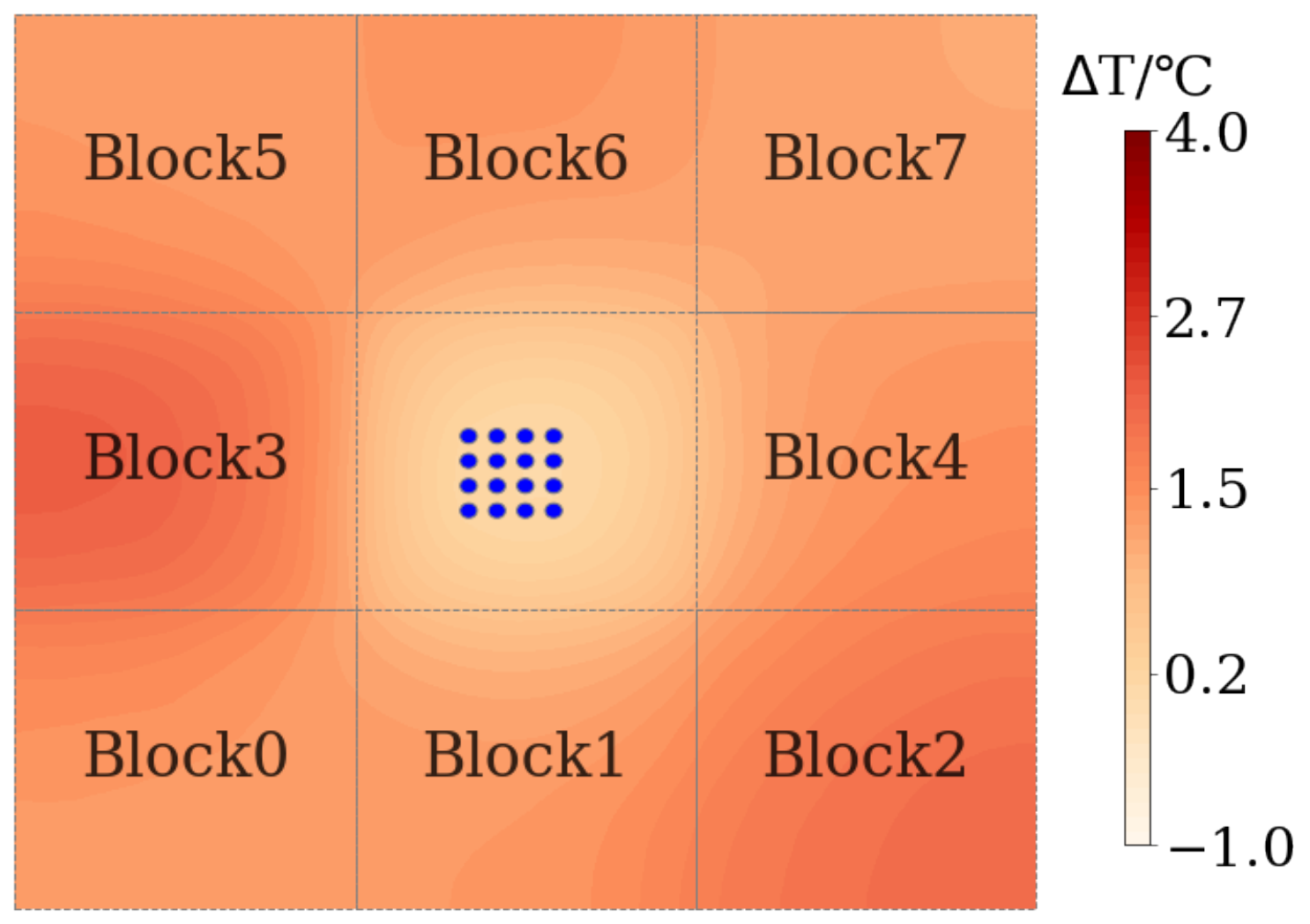}
\end{minipage}
}
\subfloat[nonlinear leakage power]{
\begin{minipage}[t]{0.48\linewidth}
    \centering
    \includegraphics[width=0.95\linewidth]{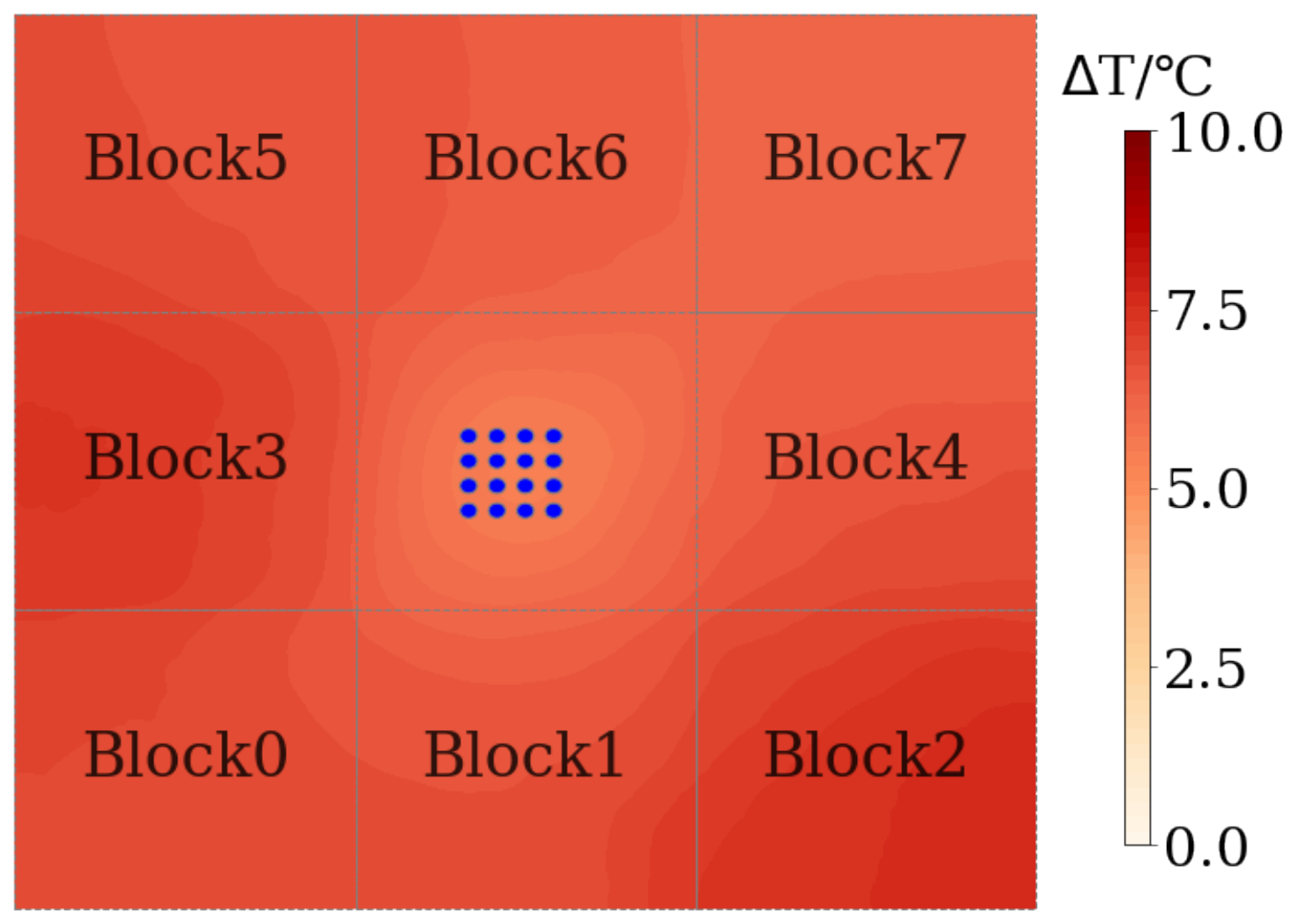}
\end{minipage}
}
\centering
\caption{
Temperature difference induced by considering the (a) nonlinear thermal conductivity of Si and (b) nonlinear leakage power in a TSV-based 3D-IC (a $4\times4$ TSV array in the middle, shown by blue circles), compared to the linear case.}
\vspace{-0.3cm}
\label{Fig:Nonlinear}
\end{figure}

The third challenge is that various nonlinear thermal mechanisms, such as nonlinear thermal conductivity and nonlinear leakage power accompany thermal hotspots and further exacerbate the high temperature in ICs. We illustrate these effects in Fig. \ref{Fig:Nonlinear}, by simulating the temperature difference induced by considering the non-linear conductivity of Si and non-linear leakage power in a TSV-based 3D-IC, compared to the case with constant conductivity and power. The temperature will increase by at most $10^\circ C$ if both mechanisms are taken into account. Although some previous efforts \cite{yan2017efficient,bagnall2013application,ladenheim2018mta} have investigated either one of the mechanisms, a comprehensive simulation tool is lacking. 

Confronted with these challenges, we propose ATSim3D\footnote{Binary \& cases available: https://github.com/Brilight/ATSim3D\_pub.git}, an accurate thermal simulator for nonlinear and heterogeneous simulation of 3D-IC systems. We summarize the capabilities of other relevant simulators in Table \ref{table:comp}, and compare them with our ATSim3D. In this way, we conclude our major contributions:
\begin{itemize}
\vspace{-1pt}
    \item We propose an efficient and accurate steady-state thermal simulator, considering  both the nonlinear thermal conductivity and leakage power in 3D-ICs.
    \item We propose a global-local approach for simulation with ultra-high resolution ($4096\times4096$ or even higher) and high efficiency by local-level parallelization.
    \item We propose utilizing the Kirchhoff transformation to address the nonlinear conductivity and prove the continuity condition for the transformed equation.
    \item We validate the ATSim3D for both 2D and 3D (Mono3D and TSV-based) ICs. It exhibits high accuracy (relative error $<3\%$, max error $<\SI{3}{\degreeCelsius}$), and efficiency ($40\times$ acceleration in average) compared to rigorous simulator COMSOL. 
\vspace{-1pt}
\end{itemize}
The rest of this paper is organized as follows.
Section~\ref{sec:prelim} reviews the basic thermal model and formulates the accuracy criteria.
Section~\ref{sec:algo} provides a thorough explanation of the proposed simulator.
Section~\ref{sec:result} demonstrates the power of ATSim3D with comprehensive results, followed by the conclusion in Section~\ref{sec:conclu}.

\begin{table}[tbh]
\footnotesize
\centering
\vspace{-0.2cm}
\caption{Comparison of thermal simulators' features.}
\resizebox{0.48\textwidth}{!}{
\begin{tabular}{|c|ccc|cc|}
\hline
        Simulator & Algorithm & Resolution & Efficiency & \makecell[c]{Nonlinear\\Leakage} & \makecell[c]{Nonlinear\\Conductivity} \\ \hline
        COMSOL [Commercial] & \multirow{2}{*}{FEM} & \textbf{High} & Low & \checkmark & \checkmark \\
        MTA \cite{ladenheim2018mta} & & Medium & \textbf{Fast} & \checkmark & \checkmark \\ \hline
        HotSpot \cite{stan2003hotspot} & \multirow{3}{*}{FDM} & Low & Medium & \checkmark & $\times$ \\
        PACT \cite{yuan2021pact} & & \textbf{High} & \textbf{Fast} & $\times$ & $\times$ \\
        3D-ICE \cite{terraneo20213d} & & Low & Medium & $\times$ & $\times$ \\ \hline
        ATSim3D (ours) & FVM & \textbf{High} & \textbf{Fast} & \checkmark & \checkmark \\ \hline
\end{tabular}
}
\vspace{-0.2cm}
\label{table:comp}
\end{table}

\section{Preliminaries} \label{sec:prelim}
In this section, we first introduce the 3D-IC configurations (\ref{sec:3dconfig}). Then the thermal models (\ref{sec:models}) are constructed, and different simulators are compared. 
Finally, we formulate the problem in \ref{sec:metric}.

\subsection{3D-IC Configurations} \label{sec:3dconfig}

Fig. \ref{Fig:3DIC} shows the simplified 3D-IC structures, including both (a) monolithic 3D-IC \cite{shukla2019overview} and (b) TSV-based 3D-IC \cite{hanhua2014thermal}. Layers in the Mono3D-IC are fabricated sequentially.
Power sources are located at the top of the upper and bottom tiers, above which the metal layers are stacked. In TSV-based 3D-ICs, the active layers contain functional blocks as power sources, 
while the interconnect layers are modeled to be homogeneous with an effective thermal conductivity since the detailed routing information is agnostic. TSVs punch from the top of tier 1 up to the top of tier 2, as shown in Fig. \ref{Fig:3DIC}(b). 

There exist two heat dissipation paths: the primary heat flow path, composed of heat spreader, heat sink, and package, and the secondary heat flow path, including package substrate and PCB. As a common practice \cite{yuan2021pact}, we simplify the primary path to dissipate heat into the ambient through convection and ignore the secondary path.
Namely, we impose the adiabatic boundary condition on each surface except the bottom, ignoring intricate mechanisms. 

\subsection{Thermal Models} \label{sec:models}

\subsubsection{Nonlinear thermal PDE} Since thermal designs typically concern steady-state temperature profiles under certain worst cases, we focus only on  steady-state thermal analysis. The governing heat diffusion equation reads:
\begin{equation}
    \nabla\cdot\left(\kappa(\textbf{r},T)\nabla T(\textbf{r})\right) = -\textbf{P}(\textbf{r},T),
    \label{Equ:Fourier}
\end{equation}
subject to the boundary conditions (B.C.):
\begin{align}
        \vec{n}\cdot\nabla T|_{\text{Lateral\&Top}} &= 0\ \text{(Adiabatic B.C.)} \label{equ:bc1} \\ 
        -\left(\kappa\vec{n}\cdot\nabla T\right)|_{\text{Bottom}} &= h(T-T_{amb})\ \text{(Convection B.C.)} , \label{equ:bc2}
\end{align}
where $T(\textbf{r})$ is the 3D temperature $[unit\ K]$ profile  over the location $\vec{r}=(x,y,z)$,  $\kappa(\textbf{r},T)$ is the temperature-dependant heterogeneous conductivity $[W/m\cdot K]$, $\textbf{P}(\textbf{r},T)=\textbf{P}_{\text{dyn}}(\textbf{r})+\textbf{P}_{\text{leak}}(\textbf{r},T)$ is the sum of dynamic and leakage power densities $[W/m^3]$. 
According to \cite{bonani1995application}, the conductivity of common materials in IC (Si, Cu, GaAs, etc.) follows a power law with the temperature: 
\begin{equation}
    \kappa(T)= \kappa_0 (T_0/T)^\alpha, \label{equ:kappa}
\end{equation}
where $\kappa_0$ is the conductivity at temperature $T_0$, and $\alpha$ is a power law constant. $\alpha\sim 1.5$ for silicon, while $\alpha\sim 0$ for copper in the typical temperature range of IC. As for the power, the dynamic power of blocks exhibits little dependence on the temperature, while leakage power shows explicit reliance, following the relation \cite{lu2017temperature}:
\begin{equation}
    \textbf{P}_{\text{leak}}= \textbf{P}_{\text{L0}}\cdot e^{\beta (T-T_0)},
\label{equ:leakage}
\end{equation}
where $\textbf{P}_{\text{L0}}$ is the base leakage power, and $\beta,\ T_0$ the temperature coefficient and base temperature. 

\subsubsection{Previous works} The nonlinear leakage and conductivity mechanisms \cite{yan2017efficient, ramalingam2006accurate} are typically tackled separately, where  Newton methods are employed. To speed up, direct methods are proposed based on certain assumptions. For nonlinear conductivity, Bonani \textit{et al.} \cite{bonani1995application} proposes to transform the Eq. \ref{Equ:Fourier} into the linear equation by the Kirchhoff transformation, based on the hypothesis that the temperature-dependent behavior for all materials is the same, which does not stand firm for TSV-based systems. In \cite{oh2012efficient}, virtual power sources are introduced to replace the TSV with homogeneous Si, which does not apply to large-scale heterogeneous systems. For nonlinear power profiles, approximate forms of leakage power, including linear model or piecewise linear model \cite{sultan2020fast}, are proposed to speed up calculation, at the cost of accuracy.

\vspace{-0.15cm}
\subsection{Problem Formulation} \label{sec:metric} 
In this work, we focus on solving the nonlinear equation Eq. \ref{Equ:Fourier} under the B.C.s Eq. \ref{equ:bc1} and \ref{equ:bc2}, with the temperature dependence of conductivity and leakage power defined in Eq. \ref{equ:leakage} and \ref{equ:kappa}. 
To measure the accuracy of the simulation results, we define three error metrics: mean absolute error ($\text{MAE} = \textbf{Mean}\{\|T_{AT}-T_{ref}\|\}$), maximum error ($\text{MaxE} = \textbf{Max}\{\|T_{AT}-T_{ref}\|\}$), and mean absolute relative error ($\text{MARE} = \textbf{Mean}\{\frac{T_{AT}-T_{ref}}{T_{ref}-T_\text{amb}}\}$), 
where $T_{AT}$, $T_{ref}$ are the temperature profile calculated by ATSim3D and rigorous simulators, and the ambient temperature $T_\text{amb}$ is introduced in \text{MARE} for regularization.
\begin{figure}[tbh]
\vspace{-0.3cm}
\centering
\includegraphics[height=8.4cm]{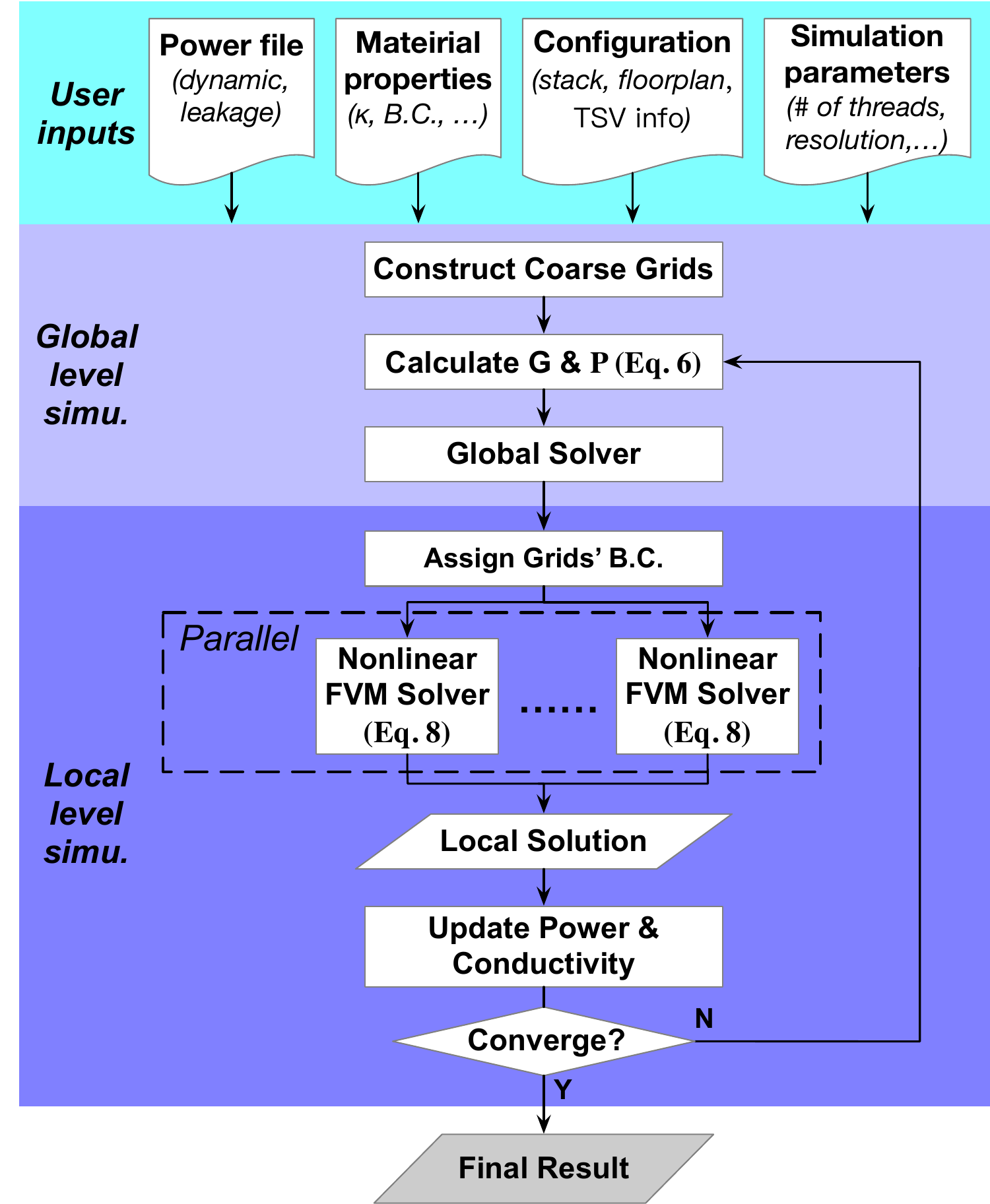}
\vspace{-0.2cm}
\caption{The ATSim3D simulation flow, consisting of iterations between the coarse level and fine level simulation.}
\label{fig:flow}
\end{figure}

\begin{figure}[tbh]
\vspace{-0.4cm}
\includegraphics[width=1.05\linewidth]{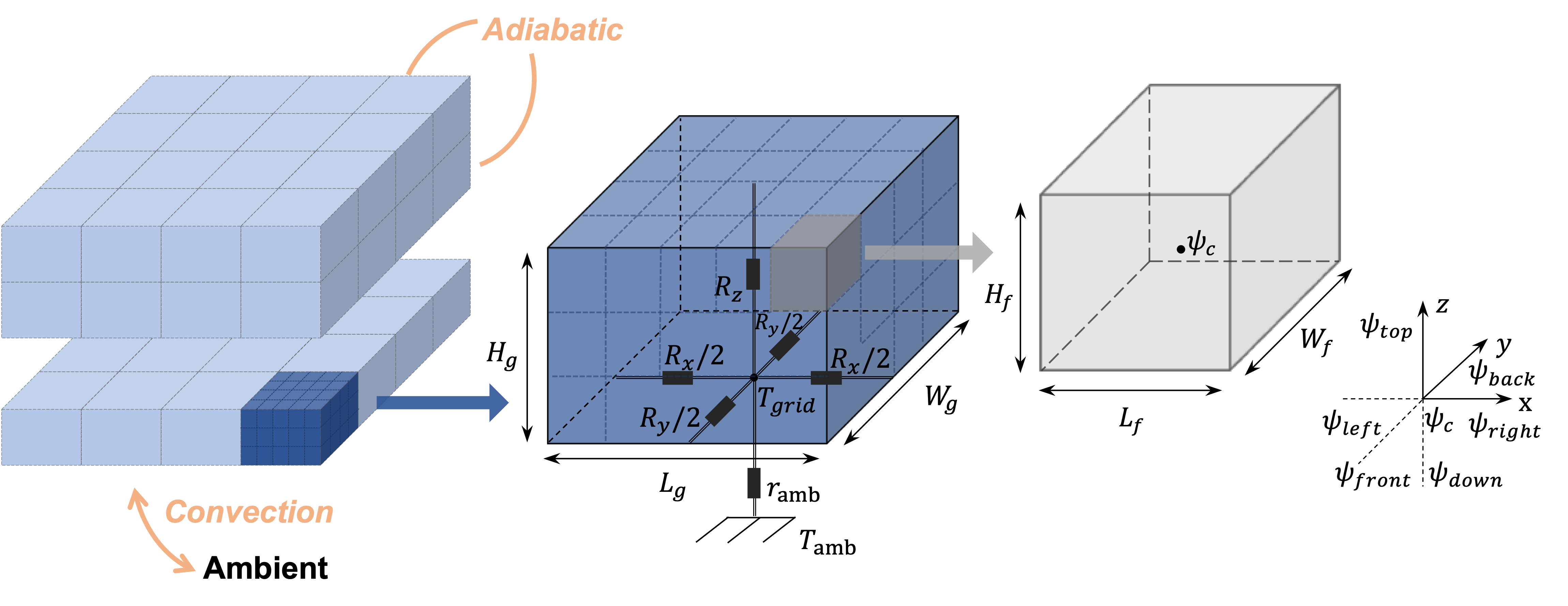}
\vspace{-0.4cm}
\caption{The hierarchical partition of the whole system. From left to right: whole package, global coarse grid, and local fine grid. The geometric dimensions and unknown variables are indicated. }
\vspace{-0.3cm}
\label{fig:partition}
\end{figure}

\section{Whole Framework} \label{sec:algo}
In this section, we illuminate the algorithm of ATSim3D. We first introduce the motivation (\ref{motiv}), then detail the global (\ref{global}) and local (\ref{local}) simulation, and end up with the whole iteration loop (\ref{iters}).
\vspace{-0.2cm}
\subsection{Motivation} \label{motiv}

Thermal simulation is a multi-scale problem since the package typically features $mm$ size, while the blocks and cells are of $\mu m$ size, and the thickness of layers may vary in the range of $0.1\sim100\ \mu m$. As a result, main-stream solvers that directly discretize the whole system will consume too much computational burden to achieve high resolution. 
To alleviate this problem, we adopt a simplified global-local approach \cite{noor1986global}. It first solves the global solution on a coarse, global mesh. Then small subdomains are partitioned and their boundary conditions are extracted from the global solution. Local solutions can be calculated afterward in all subdomains in parallel. The Eq. \ref{Equ:Fourier} is naturally suitable for applying this global-local approach since the local temperature depends mainly on local properties, aka. the localized heating effect in \cite{wen2020dnn}.

What's more, we exploit the observation that the temperature is jointly determined by two components: one that generates heat, and another dissipates. Local fine-level simulation helps delve into active layers that generate heat where we care most. In contrast, the simulation of the heat dissipation components considers the macro scale information, which can be satisfied by the global simulation. Based on these considerations, we propose ATSim3D, featuring a global-local method as illustrated in Fig. \ref{fig:flow}. It conducts global level simulation with compact thermal models, and then local level parallel simulation with finite volume method. By iteration, the detailed power and conductivity distribution are updated after each step, and the iteration stops when convergence is reached.

\vspace{-0.2cm}
\subsection{Global Simulation}\label{global}

The principle behind the global simulation is the duality between the thermal resistance network and the electrical counterpart \cite{stan2003hotspot}. 
Given the geometrical configuration and material parameters, we first construct coarse uniform grids for the whole system, as shown in Fig. \ref{fig:partition}. For simplicity, we assume a system of cubic domain, while the method can be generalized to other configurations. In Fig. \ref{fig:partition}, each coarse grid is associated with thermal resistances in the x,y, and z directions, connecting to other coarse grids or the ambient. For a grid composed of uniform material, the resistances are:
\begin{equation}
R_{x}={L_g}/(\kappa H_g W_g), R_{y}={W_g}/(\kappa H_g L_g), R_{z}={H_g}/(\kappa L_g W_g). \label{equ:Rxyz}    
\end{equation}
The geometry size $L_g, W_g, H_g$ are attached in the figure. However, the case gets complicated if the coarse grid is composed of heterogeneous material or material with nonlinear conductivity. 
Previous methods often calculate the effective conductivity in a volume-averaged manner \cite{6241575}, which will reduce to a simple average for the composite conductivity case (TSV array, for example). 
Instead, we employ the formulation as in \cite{nie2023efficient}. Taking the x direction as an example, the fine grid array inside the coarse grid, as shown in Fig. \ref{fig:partition}, is viewed as a connection of parallel lines. Each line represents a series of resistances in the x-direction, allowing the presence of non-uniform materials. In this way, the effective resistance is:
\begin{equation}
    R_x = L_f/\left(W_f H_f \cdot\Sigma_{j,k}({1}/{\Sigma_{i}\kappa^{-1}_{i,j,k}})\right),
\end{equation}
where $L_f, W_f, H_f$ are the size of each fine grid, and $\kappa_{i,j,k}$ represents the value of conductivity of the fine grids, indexed in the order of x,y,z. The expressions of $R_y,R_z$ can be derived in the same manner, and they reduce to  Eq. \ref{equ:Rxyz} in the homogeneous case.

Now we can set up the matrix equation for global simulation: 
\begin{equation}
    \textbf{G}\cdot \textbf{T}_g=\textbf{P}, \label{equ:global}
\end{equation}
here $\textbf{G}$ is the conductivity matrix composed of effective resistances, the elements in vector $\textbf{T}_g, \textbf{P}$ are the global temperature distribution at the bottom surface, and the total power in each coarse grid. The spatial resolution ranges in $50\sim1000\ \mu\text{m}$ in this stage. 
\vspace{-0.2cm}
\subsection{Local Simulation}\label{local}
The global simulation simulates the coarse-level heat generation and dissipation process but dismisses the details inside coarse grids. So we conduct local simulation afterward to derive detailed results, especially in the active layer(s), and simulate the nonlinear thermal effects.
In this step, we discretize the coarse grids of interest into fine grid arrays (as mentioned above, see Fig. \ref{fig:partition}). The detailed boundary conditions of the fine grids are calculated by linearly interpolating the global solution. A complete global-local approach involves iterative refinements of the interface conditions between the global and local solutions \cite{el2023asynchronous}. Although ATSim3D does not adhere to this principle as iterations are time-consuming, the solutions obtained are still accurate empirically. In the following, we will show how to solve the sub-problems inside each coarse grid in parallel. The sub-problem is to solve the nonlinear equation (Eq. \ref{Equ:Fourier}) with the Dirichlet B.C. calculated from the global solution.
\subsubsection{Kirchhoff transformation}
To tackle the nonlinear equation, we impose Kirchhoff transformation to the temperature in Eq. \ref{Equ:Fourier}:
\begin{equation}
    \psi = \frac{1}{\kappa_b} \int_{T_g}^{T} \kappa(\eta) d\eta,
    \label{equ:Kirchhoff}
\end{equation}
where $\kappa_b=\kappa(T_g)$ (Eq. \ref{equ:kappa}) and $T_g$ is the base temperature of the coarse grid, derived from the global solution. For constant conductivity $\psi = T-T_g$, and for the conductivity following Eq. \ref{equ:kappa}, $\psi = \frac{T_0^\alpha}{1-\alpha}(T^{1-\alpha}-T_g^{1-\alpha})\ (\alpha\neq1)$. Eq. \ref{Equ:Fourier} is then transformed to be (given detailed power distribution in each iteration):
\begin{equation}
\nabla\cdot\left(\kappa_b(\textbf{r})\nabla\psi(\textbf{r})\right) = -\textbf{P}(\textbf{r}). \label{Equ:transformed}
\end{equation}
For homogeneous systems, the equation reduces to solving $\nabla^2\psi=-\textbf{P}(\textbf{r})/\kappa_b$. However, for heterogeneous systems, we have to pay extra attention to the continuity conditions at material interfaces. 

\subsubsection{Continuity Conditions} 
Let us consider two regions ($I,\ II$) separated by interface $\Sigma$. The continuity conditions require both the temperature and heat flow to be continuous across the interface: 
\begin{equation}
    T_I|_{\Sigma}=T_{II}|_{\Sigma},\ \ -\ \kappa(T)_I\nabla T_I|_{\Sigma}=-\kappa(T)_{II}\nabla T_{II}|_{\Sigma}.
\end{equation}
The second condition still holds for the transformed variable, since $\kappa_{b,\ i}\nabla \psi_{i}\equiv\kappa(T)_{i}\nabla T_{i}\ (i=I,II)$. 
Conversely, the first condition does not hold at first glance, as the relation between $\psi$ and temperature varies between materials. In \cite{bonani1995application}, the authors show that to alleviate the jump of $\psi$ at the interface, the $\kappa(T)_{I}/\kappa(T)_{II}$ must be a constant, which means the temperature dependence of the conductivity of all materials must be the same. However, this assumption does not hold in most cases, especially for the TSV-based 3D-ICs, composed of copper (constant conductivity) and silicon ($\alpha=1.5$). 
Confronted with this problem, we prove that the first continuity condition will hold approximately in our global-local approach.

\vspace{-0.3cm}
\begin{proof}
Consider the interface of copper and silicon, the temperature ($T_{\text{Si}},T_{\text{Cu}}$) around the interface $\Sigma$ satisfies $|T_{\text{Si}}|_{\Sigma}-T_g|<<T_g (T_g$ in Eq. \ref{equ:global}), since the detailed temperature profile inside will not vary dramatically. Then the transformed variable satisfies:
\begin{align} 
    \tiny\psi_{\text{Si}}|_{\Sigma}
    &\tiny=\frac{T_0^\alpha T_g^{1-\alpha}}{1-\alpha}[(1+\frac{T_{\text{Si}}|_{\Sigma}-T_g}{T_g})^{1-\alpha}-1] \\
    \textit{\small(To first order)}&\approx\frac{T_0^\alpha T_g^{1-\alpha}}{1-\alpha}[1+(1-\alpha)\frac{T_{\text{Si}}|_{\Sigma}-T_g}{T_g}-1]\\
    &\tiny=(\frac{T_0}{T_g})^{\alpha}(T_{\text{Si}}|_{\Sigma}-T_g).
\end{align}
When further choose $T_0=T_g (T_0$ in Eq. \ref{equ:kappa}) in the global simulation, it concludes that $\psi_{\text{Si}}|_{\Sigma}\approx T_{\text{Si}}|_{\Sigma}-T_g=T_{\text{Cu}}|_{\Sigma}-T_g=\psi_{\text{Cu}}|_{\Sigma}$.
\end{proof}
\vspace{-0.1cm}
Now the continuity conditions can be satisfied for the variable $\psi$ at interfaces, and the Eq. \ref{Equ:transformed} inside the whole heterogeneous system can be discretized and solved by FVM.

\subsubsection{Finite Volume Method} 
For each coarse grid, we assign the variable $\psi_{i,j,k}$ to the cubic center of each fine grid, as shown in Fig. \ref{fig:partition}. FVM integrates the Eq. \ref{Equ:transformed} over the fine grid domain $\Omega$:
\begin{equation}
    \Sigma_{\gamma=x,y,z} \int_{\Omega}\partial_\gamma(\kappa_b\partial_\gamma\psi_{i,j,k}) \text{dxdydz} = -\textbf{P}_{i,j,k}L_cW_cH_c. \label{equ:integral}
\end{equation}
Integrating the x-direction component of the left term we can derive $[(\kappa_b\partial_x\psi)_{i+0.5,j,k}-(\kappa_b\partial_x\psi)_{i-0.5,j,k}]W_cH_c$. For the fine grid not at the array boundary, the expression can be transformed as:
\begin{equation*}
    [\Tilde{\kappa}_{i+0.5,j,k}(\psi_{i+1,j,k}-\psi_{i,j,k})-\Tilde{\kappa}_{i-0.5,j,k}(\psi_{i,j,k}-\psi_{i-1,j,k})]{W_cH_c}/{L_c}.
\end{equation*} 
While for the fine grid just at the boundary, like $i=0$, the gradient term is calculated according to the Dirichlet B.C.:
\begin{equation*}
    [\Tilde{\kappa}_{0.5,j,k}(\psi_{1,j,k}-\psi_{0,j,k})-2\times{\kappa}_{0,j,k}(\psi_{0,j,k}-\psi_{j,k}|_{\text{BC}})]{W_cH_c}/{L_c}.
\end{equation*} 
Here $\Tilde{\kappa}_{i\pm0.5,j,k}$ represents the effective conductivity at the interface of fine grid $(i,j,k)$ and $(i+1,j,k)$. Rather than simply calculating the mean of $\kappa_{i,j,k}$ and $\kappa_{i+1,j,k}$, we instead use their harmonic average to keep the heat flow conservative across the interface \cite{thomas2013numerical}: 
\begin{equation}
    \Tilde{\kappa}_{i+0.5,j,k} = {2}/{(\kappa_{i,j,k}^{-1}+\kappa_{i+1,j,k}^{-1})}.
\end{equation}

Summing over all three directions in Eq. \ref{equ:integral}, and incorporating all boundary values, the nonlinear equation Eq. \ref{Equ:Fourier} now reduces to a linear matrix equation of $\psi_{i,j,k}$. Local temperature results can be derived by the inverse Kirchhoff transformation easily from $\psi$. 

\subsubsection{Parallelization} 
In the local simulation, only the coarse grids that generate heat or feature nonlinear thermal effects will be simulated in detail. The sub-problems inside each coarse grid are independent after the B.C. is calculated from the global solution. Consequently, the local simulation can be conducted in a highly parallel manner for each coarse gird. Combining the results from the global and local simulation, we can derive the multi-scale temperature distribution of the whole system at the end of each global-local simulation step.

\subsection{Iteration Loop} \label{iters}
For systems that do not consider nonlinear thermal mechanisms, simply applying the aforementioned global-local simulation once is sufficient to derive an accurate result. However, iterations are necessary when nonlinear mechanisms are considered. As shown in Fig. \ref{fig:flow}, starting from a preset initial temperature (the ambient temperature or higher), we calculate the conductivity and leakage power distribution, based on which the global solution $T_g$ and local solution are calculated sequentially. Then we update the leakage power within each block/unit according to the detailed local temperature profile. Different forms of leakage power are supported in this framework, and we choose Eq. \ref{equ:leakage} for all blocks for simplicity. 

Subsequently follows another new iteration step. It first updates the conductivity distribution and corresponding thermal resistances and then conducts the global-local simulation.
The loop is terminated when the difference between the temperature distribution of the current and the previous step is small enough, typically less than $0.1^{\circ}C$. In summation, by employing iterative refinement, we can deduce the multi-scale temperature profile of non-linear and heterogeneous 3D-ICs. 
\section{Experimental Results} \label{sec:result}

\subsection{Experimental Setup}\label{sec:setup}
We arrange experiments for three IC structures, including a 2D Intel Multi-core CPU (labeled as \textbf{2DIC}) \cite{yuan2021pact}, a 7-layer complex Mono3D-IC with 2 active layers (\textbf{Mono3D}) \cite{shukla2019overview}, and a two-tier TSV-based 3D-IC (\textbf{TSV3D}) \cite{hanhua2014thermal}), and four  configurations of parameters (based on whether two nonlinear effects are considered or not, categorized in Table \ref{table:config}). The floorplan and power maps are adopted from relevant works. As shown in Table \ref{table:atfig}, the \textbf{Mono3D} features a random power map with 2500 units, and there exists a $4\times4$ TSV array (cylinder shape, radius $15\mu m$, pitch $50\mu m$) in the middle of \textbf{TSV3D}, with an oversized keep-out-zone. 
ATSim3D is written in Python, and the parallel computing is realized by the \textit{multiprocessing.Pool} module, allowing the tasks to be offloaded to multiple cores. We perform the simulation on a Linux server with Intel Xeon 2.10 GHz processors with a maximum of 64 cores and 128 GB memory.

\begin{table}[tbh]
\vspace{-0.2cm}
\caption{The parameter configurations used in the experiments, concerning the nonlinear conductivity and leakage power.}
\centering
\normalsize
\resizebox{0.47\textwidth}{!}{
\begin{tabular}{c|cc}
\toprule
    Ambient $40^{\circ}C$ &$\kappa_{Si}=150\ W/(m\cdot K)$
    &$\kappa_{Si}=153\cdot(300K/T)^{1.5} $\\
    \midrule
    Constant Leakage @ $85^{\circ}C$ & I & II \\
    $\beta$=0.015, $T_0 =85^{\circ}C$ (Eq. \ref{equ:leakage}) & III & IV \\
\bottomrule
\end{tabular} 
}
\vspace{-0.2cm}
\label{table:config}
\end{table}

\subsection{Accuracy Analysis}\label{sec:accu}

We first validate the accuracy of ATSim3D against COMSOL. The error metrics defined in Section \ref{sec:metric} are calculated for the z-axis central cross of the top active layer and shown in Table \ref{table:result-linear} and \ref{table:result}. 
The global and local grid resolutions are labeled by $Reso_{global}^2, Reso_{local}^2$ in the \textbf{Reso} columns. Local grid resolution represents the total number of fine grids in each layer. 
In Table \ref{table:result-linear}, we first compare the results between three typical numerical methods in Table \ref{table:comp}: COMSOL (FEM), PACT (FDM), and ATSim3D (FVM) on configuration \textbf{I}. ATSim3D can achieve relative errors <3\%, comparable to PACT. Note that PACT cannot handle TSV and it also fails on \textbf{2DIC-I} with $2048 \times 2048$ grids and \textbf{Mono3D-I} with $800 \times 800$ grids after using up all the memory. In contrast, ATSim3D is memory-efficient.
In Table \ref{table:result}, we compare COMSOL and ATSim3D on configurations \textbf{II-IV}. We do not compare with PACT as it can not support nonlinear configurations. We can see that ATSim3D shows a mean absolute error of $<1^{\circ}C$, a maximum error of $3^{\circ}C$, and a relative error of $<3\%$.
Additionally, the nonlinear thermal effects are reflected by the difference between maximum temperature values of \textbf{TSV3D-IV} and \textbf{TSV3D-I}, as high as $10^{\circ}C$. 

Simulation results of ATSim3D and COMSOL and the error distribution are provided in Table \ref{table:atfig}. We can observe a close match between the simulated results, while the error concentrates in the area of high temperature or close to the boundary, a result stemming from the different ways of mesh (grid) partitioning.

\begin{figure}[tbh]
\vspace{-0.3cm}
\centering
\subfloat[Error --- number of iterations]{
\begin{minipage}[t]{0.48\linewidth}
    \centering
    \includegraphics[width=1\linewidth]{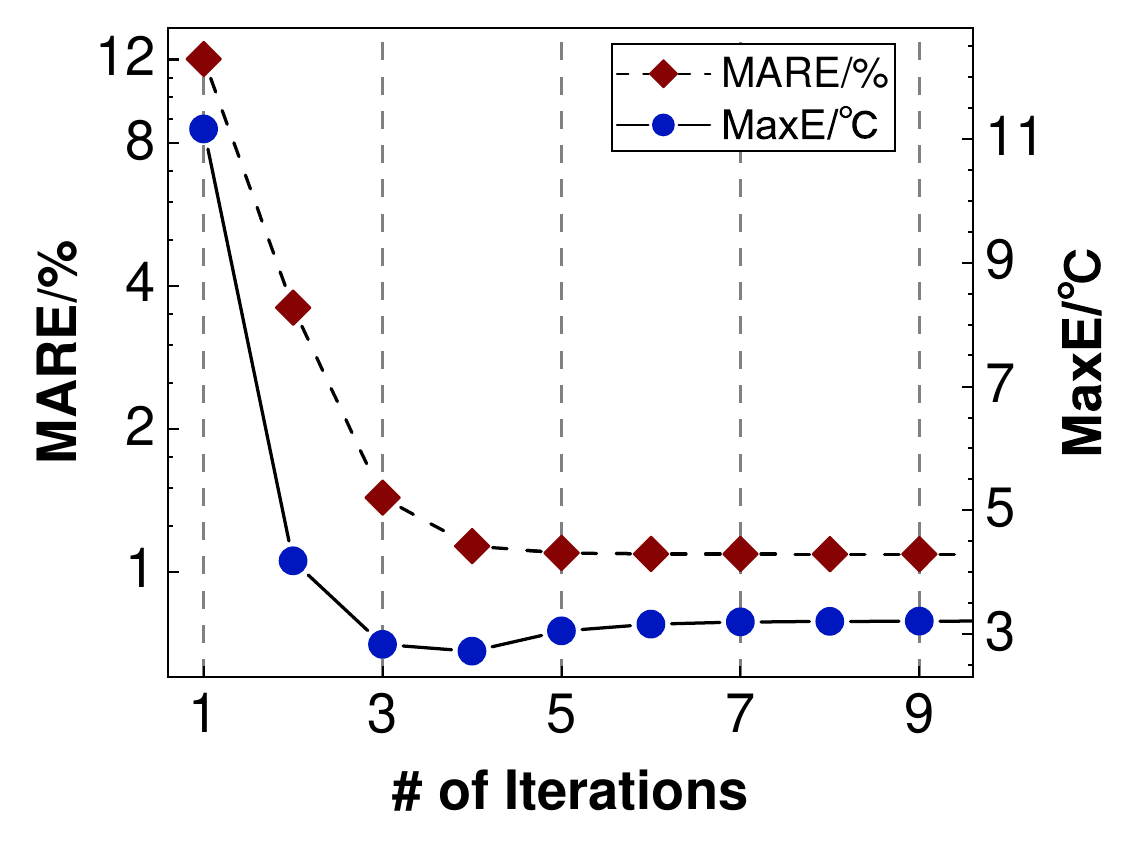}
\end{minipage}
}
\subfloat[Error --- global grid resolutions]{
\begin{minipage}[t]{0.49\linewidth}
    \centering
    \includegraphics[width=1\linewidth]{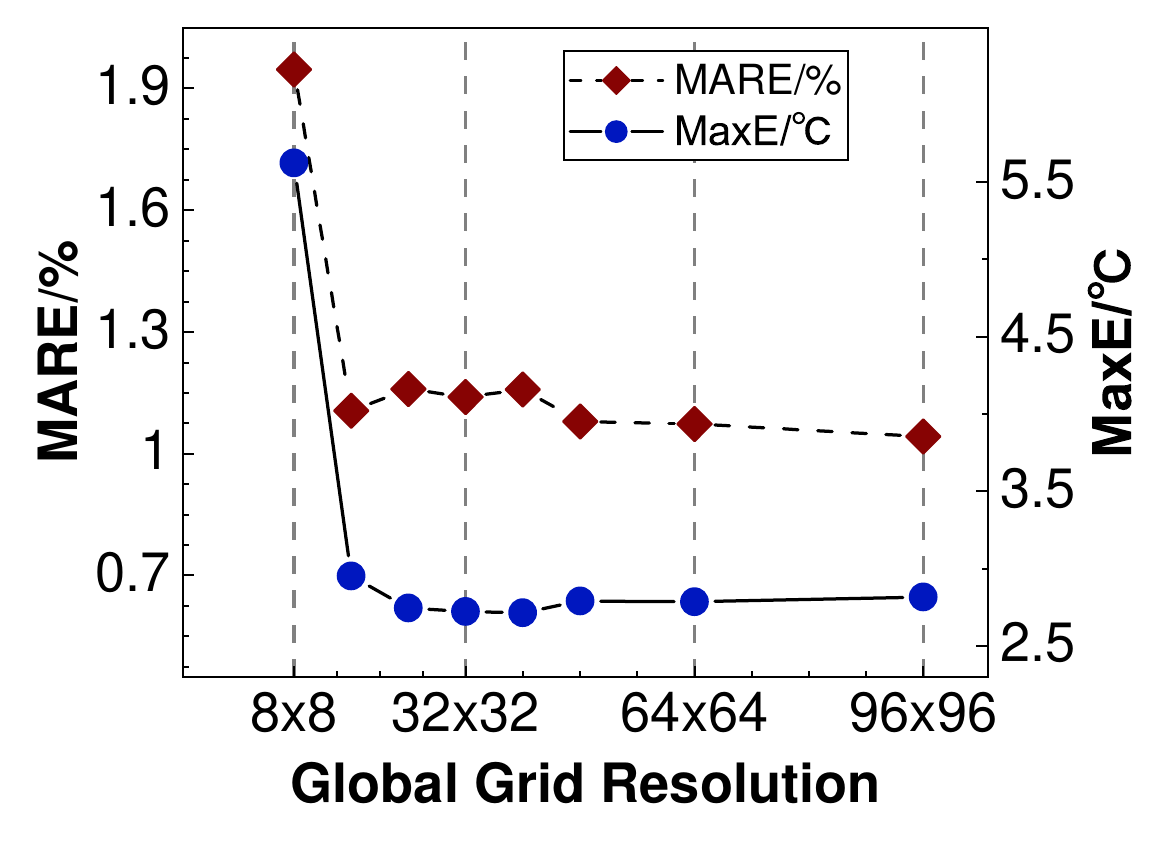}
\end{minipage}
}
\centering
\caption{The relation between the error of ATSim3D in \textbf{TSV3D-IV} and (a) the number of iterations, as well as  (b) the global grid resolutions, with each coarse grid partitioned into $16\times16$ fine grids. }\label{Fig:error}
\vspace{-0.3cm}
\end{figure}

Fig. \ref{Fig:error} provides further analysis of the factors affecting the error in \textbf{TSV3D-IV}. We can see that the error first decreases then saturates, as the number of iterations (Fig. \ref{Fig:error}(a)) and global grid resolution (Fig. \ref{Fig:error}(b)) increase. ATSim3D requires just 3-4 iterations to get converged, and it gets a mesh-independent solution at the grid resolution of $\sim6\mu m$ as the power map is coarse. Higher resolution of $\sim1\mu m$ is achieved for the \textbf{Mono3D} in Table \ref{table:result-linear}. To further reduce the error, attention should be paid to polishing the B.C. in the local simulation \cite{gosselet2018non}, which lacks some high frequency and local information since it is calculated by interpolating the global solution.

\begin{table}[tbh]
\vspace{-0.2cm}
\caption{Performance comparison between COMSOL, PACT, and ATSim3D for the linear configuration \textbf{I}. \textbf{Reso} and \textbf{Reso}$_{local}$ represent the total number of fine grids in the active layers for PACT and ATSim3D, respectively.}
\centering
\normalsize
\resizebox{0.5\textwidth}{!}{
\begin{tabular}{c|ccc|cccc|cccc}
\toprule
    \multirow{2}{*}{\textbf{Structure}} 
    & \multicolumn{3}{c|}{\textbf{COMSOL}}
    & \multicolumn{4}{c|}{\textbf{PACT}} 
    & \multicolumn{4}{c}{\textbf{ATSim3D}}  \\
    & $T_{max}/\SI{ }{\degreeCelsius}$ &Memory & Time/s 
    &Reso &Memory &Time/s & MARE/\%
    & Reso$_{local}$ &Memory &Time/s &MARE/\%  \\ \midrule
    \multirow{3}{*}{\textbf{2DIC-I}} 
    & \multirow{5}{*}{69.32} & \multirow{5}{*}{3.0 GB} & \multirow{5}{*}{60}
    & 256$^2$ & 590 MB & 8 & 2.38 & 256$^2$ & 121 MB & 6 & 2.41 \\
    & & & & 512$^2$ & 2.5 GB & 55 & 2.40 & 512$^2$ & 209 MB & 8 & 2.07 \\
    & & & & 1024$^2$ & 11.9 GB & 450 & 2.40 & 1024$^2$ & 419 MB & 11 & 2.09 \\
    & & & & 2048$^2$ & \multicolumn{3}{c|}{\textit{Not enough memory}} & 2048$^2$ & 1.1 GB & 34 & 2.09 \\
    & & & & 4096$^2$ & \multicolumn{3}{c|}{\textit{Not enough memory}} & 4096$^2$ & 4.3 GB & 94 & 2.09 \\ \midrule
   \multirow{2}{*}{\textbf{Mono3D-I}} 
   & \multirow{3}{*}{89.61} & \multirow{3}{*}{12.9 GB} &\multirow{3}{*}{1410} 
    & 200$^2$ & 6.5 GB & 255 & 0.53 & 200$^2$ & 412 MB & 15 & 0.55 \\
    & & & & 400$^2$ & 39.2 GB & 2980 & 0.53 &400$^2$ & 434 MB & 18 & 0.54 \\
    & & & & 800$^2$ & \multicolumn{3}{c|}{\textit{Not enough memory}} &800$^2$ & 536 MB & 23 & 0.54 \\ \midrule
   \textbf{TSV3D-I} & 119.58 & 5.5 GB & 77 
   & \multicolumn{4}{c|}{\textit{Unable to handle TSV}}  &640$^2$ & 337 MB & 10 & 0.14 \\
\bottomrule
\end{tabular} 
} 
\label{table:result-linear}
\end{table}

\begin{table}[tbh]
\vspace{-0.3cm}
\caption{Performance comparison between ATSim3D and COMSOL for the nonlinear configurations (\textbf{II-IV}).}
\centering
\normalsize
\resizebox{0.5\textwidth}{!}{
\begin{tabular}{c|cc|ccc|ccc}
\toprule
    \multirow{2}{*}{\textbf{Configuration}} & \multicolumn{2}{c|}{\textbf{COMSOL}} & \multicolumn{3}{c|}{\textbf{ATSim3D}} & \multicolumn{3}{c}{\textbf{Error Metric}} \\
    & $T_{max}/\SI{ }{\degreeCelsius}$ & Time/s & Time/s & Reso$_{global}$ & Reso$_{local}$ & MARE/\% & MaxE/$\SI{ }{\degreeCelsius}$ & MAE/$\SI{ }{\degreeCelsius}$  \\ 
    \midrule
    {\textbf{2DIC-II}} & 69.86 & 110 & 18 & \multirow{3}{*}{64$^2$} & \multirow{3}{*}{512$^2$} & 1.45 & 1.02 & 0.12 \\
    {\textbf{2DIC-III}} & 67.13 & 101 & 14 & & & 1.47 & 0.86 & 0.12 \\
    {\textbf{2DIC-IV}} & 67.29 & 117 & 25 & & & 2.69 & 1.54 & 0.18 \\ \midrule
    
    {\textbf{Mono3D-II}} & 91.40 & 2930 & 37 & \multirow{3}{*}{50$^2$} & \multirow{3}{*}{400$^2$}  & 0.69 & 0.96 & 0.28\\
    {\textbf{Mono3D-III}} & 89.07 & 3966 & 25 & & & 0.90 & 0.58 & 0.13\\
    {\textbf{Mono3D-IV}} & 90.60 & 5541 & 53 & & & 1.06 & 1.15 & 0.41\\ \midrule

    {\textbf{TSV3D-II}} & 121.78 & 118 & 30 & \multirow{3}{*}{40$^2$} & \multirow{3}{*}{640$^2$} & 0.30 & 1.26 & 0.21 \\
    {\textbf{TSV3D-III}} & 126.96 & 106 & 25 & & & 0.65 & 1.40 & 0.49 \\
    {\textbf{TSV3D-IV}} & 130.16 & 109 & 39 & & & 1.15 & 2.71 & 0.92  \\
\bottomrule
\end{tabular} 
} 
\vspace{-0.3cm}
\label{table:result}
\end{table}

\begin{figure}[h]
\vspace{-0.3cm}
\centering
\subfloat[Runtime --- number of cores]{
\begin{minipage}[t]{0.43\linewidth}
    \centering
    \includegraphics[width=1\linewidth]{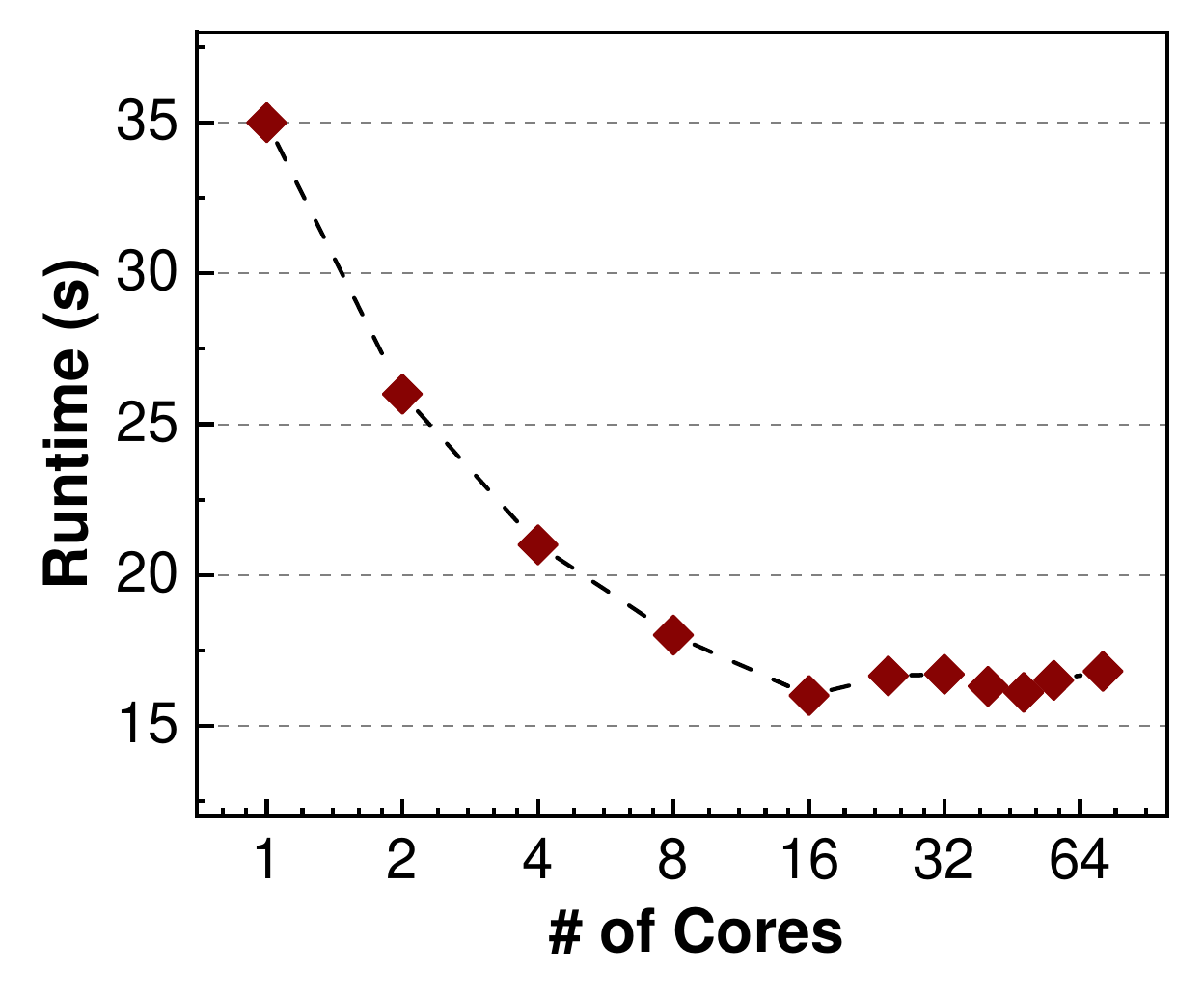}
\end{minipage}
}
\subfloat[Runtime --- grid resolutions]{
\begin{minipage}[t]{0.5\linewidth}
    \centering
    \includegraphics[width=1\linewidth]{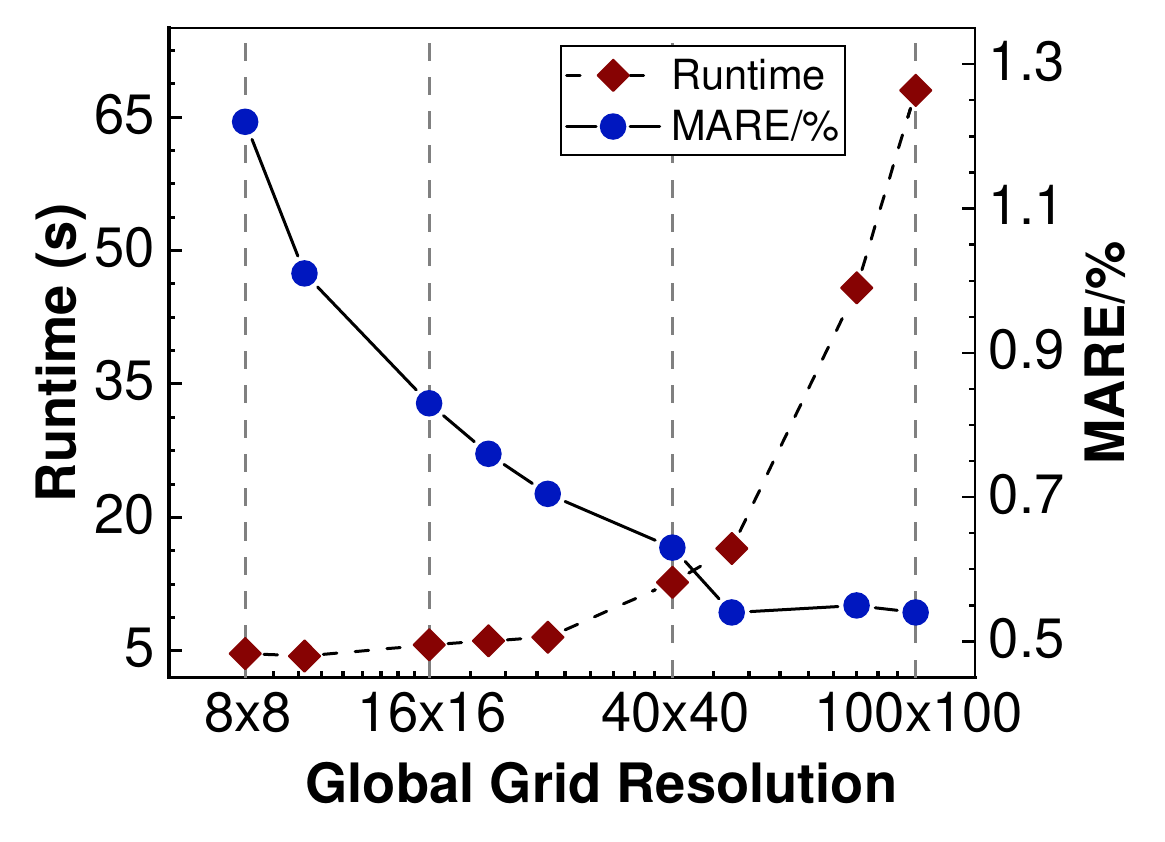}
\end{minipage}
}
\centering
\caption{The relation between the runtime of ATSim3D in \textbf{Mono3D-I} and (a) the number of cores, as well as (b) the grid resolutions, with the local resolution constant to be $400^2$.}\label{Fig:runtime}
\vspace{-0.4cm}
\end{figure}

\subsection{Efficiency Analysis}\label{sec:eff}
In this section, we analyze the efficiency of ATSim3D. Firstly, the runtime is compared in Table \ref{table:result-linear} and \ref{table:result} w.r.t COMSOL and PACT, with a maximum of 16 CPU cores used. ATSim3D achieves a speedup of $3\sim150\times$ ($40\times$ in average) compared to COMSOL, benefiting from two factors: the simplified global-local algorithm and the parallelization utilized in the local simulation. COMSOL consumes much time for the \textbf{Mono3D} due to the requirement for fine-resolution mesh division. ATSim3D is also faster than PACT (at most $40\times$), and overcomes the scalability challenge of PACT, achieving a SOTA resolution of $4096\times4096$ in the active layer.

\begin{table*}[tbh]
\vspace{-.4cm}
\caption{Results of ATSim3D and COMSOL for the nonlinear configuration in the Table \ref{table:result}, and corresponding error distribution. }
\tiny
\centering
\resizebox{1\textwidth}{!}{
\begin{tabular}{c|c|ccc}
\toprule
    Structure & Power Map & COMSOL & ATSim3D & Error ($\tiny{T_{AT}-T_{comsol}}$)\\
    \midrule
    \textbf{2DIC-IV}  
    & {\raisebox{-.4\height}{\includegraphics[height=0.5in]{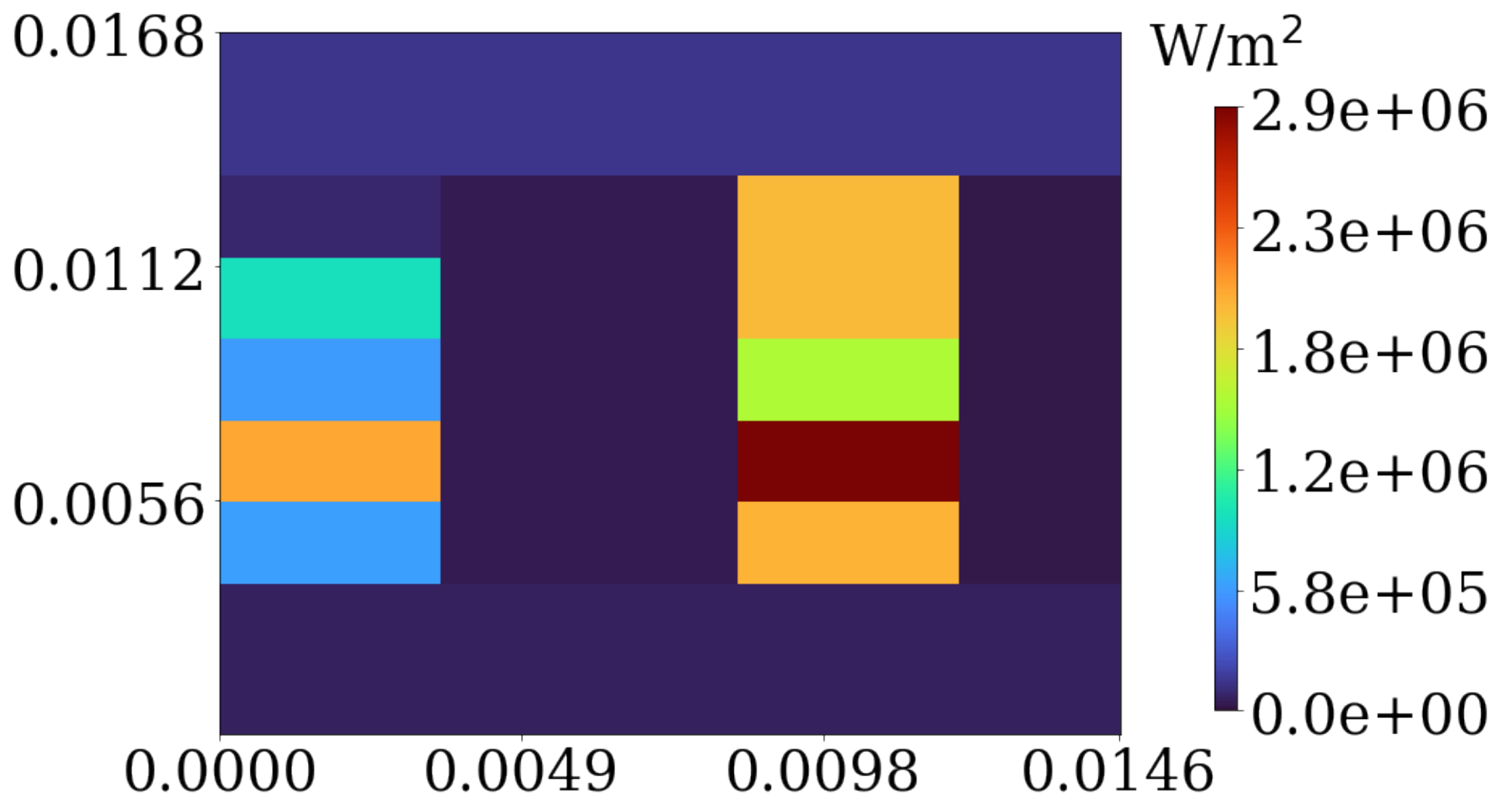}}} 
    & {\raisebox{-.4\height}{\includegraphics[height=0.5in]{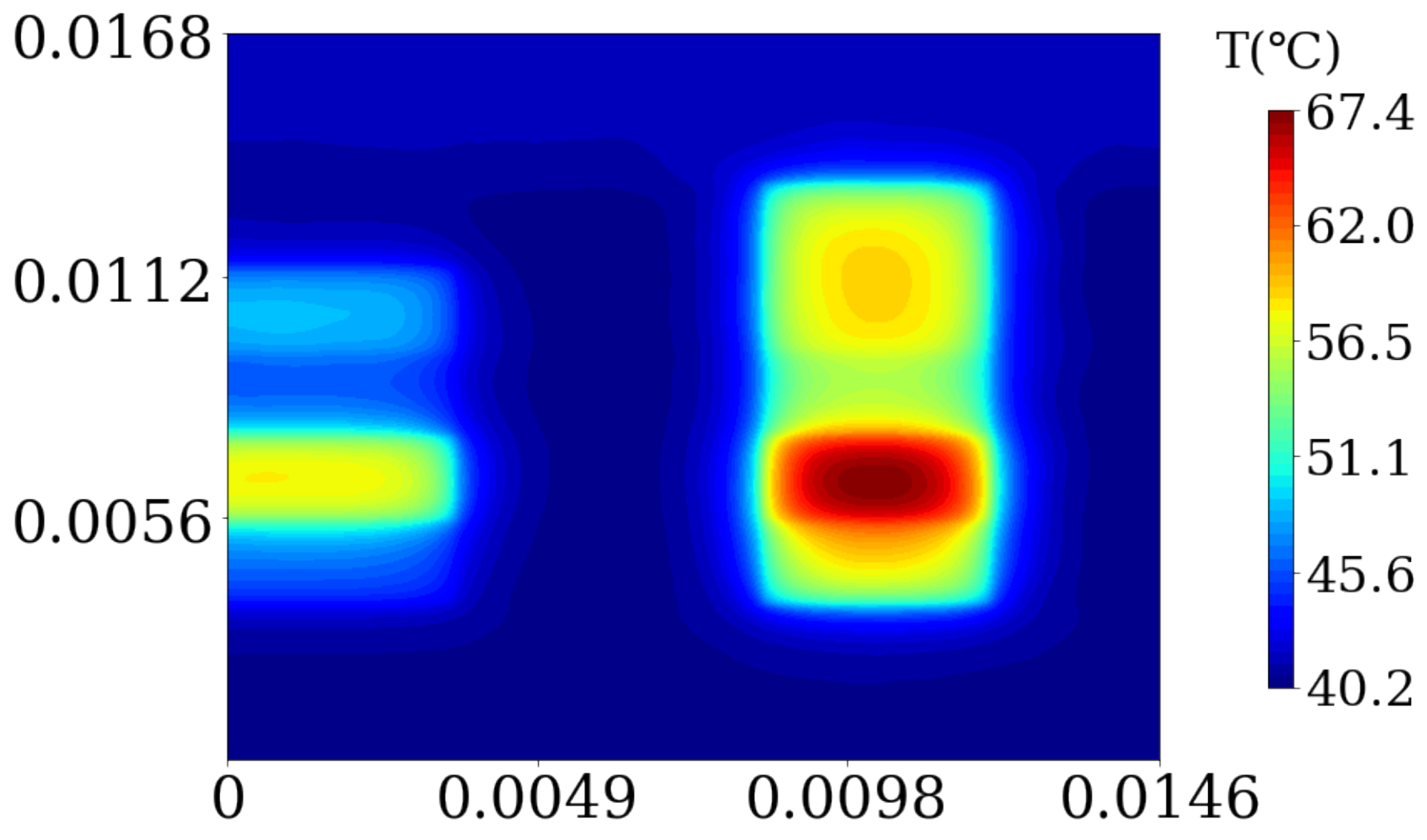}}}
    & {\raisebox{-.4\height}{\includegraphics[height=0.5in]{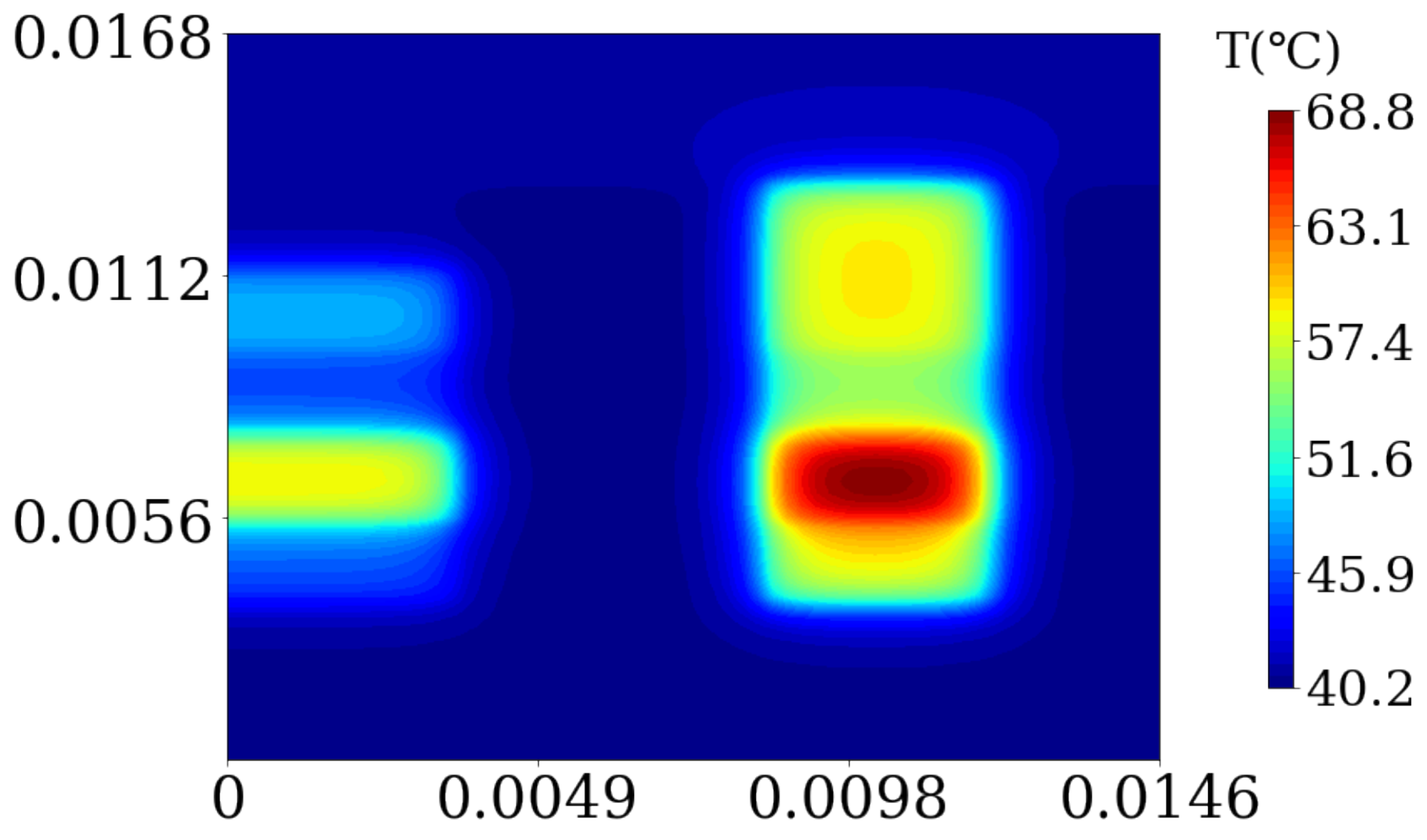}}}
    & {\raisebox{-.4\height}{\includegraphics[height=0.5in]{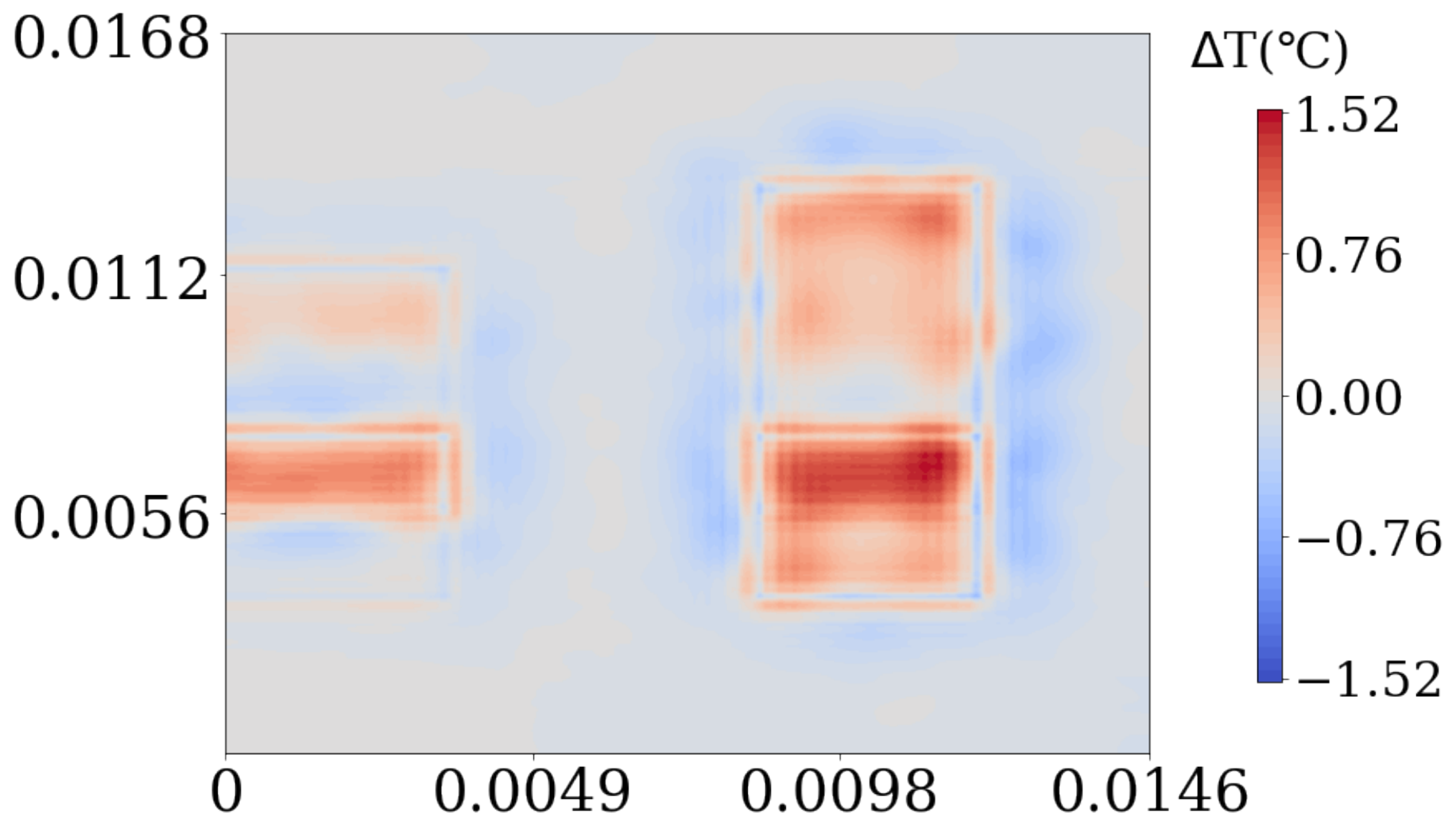}}}\\ [0.5cm] \midrule
    \textbf{Mono3D-IV}  
    & {\raisebox{-.4\height}{\includegraphics[height=0.5in]{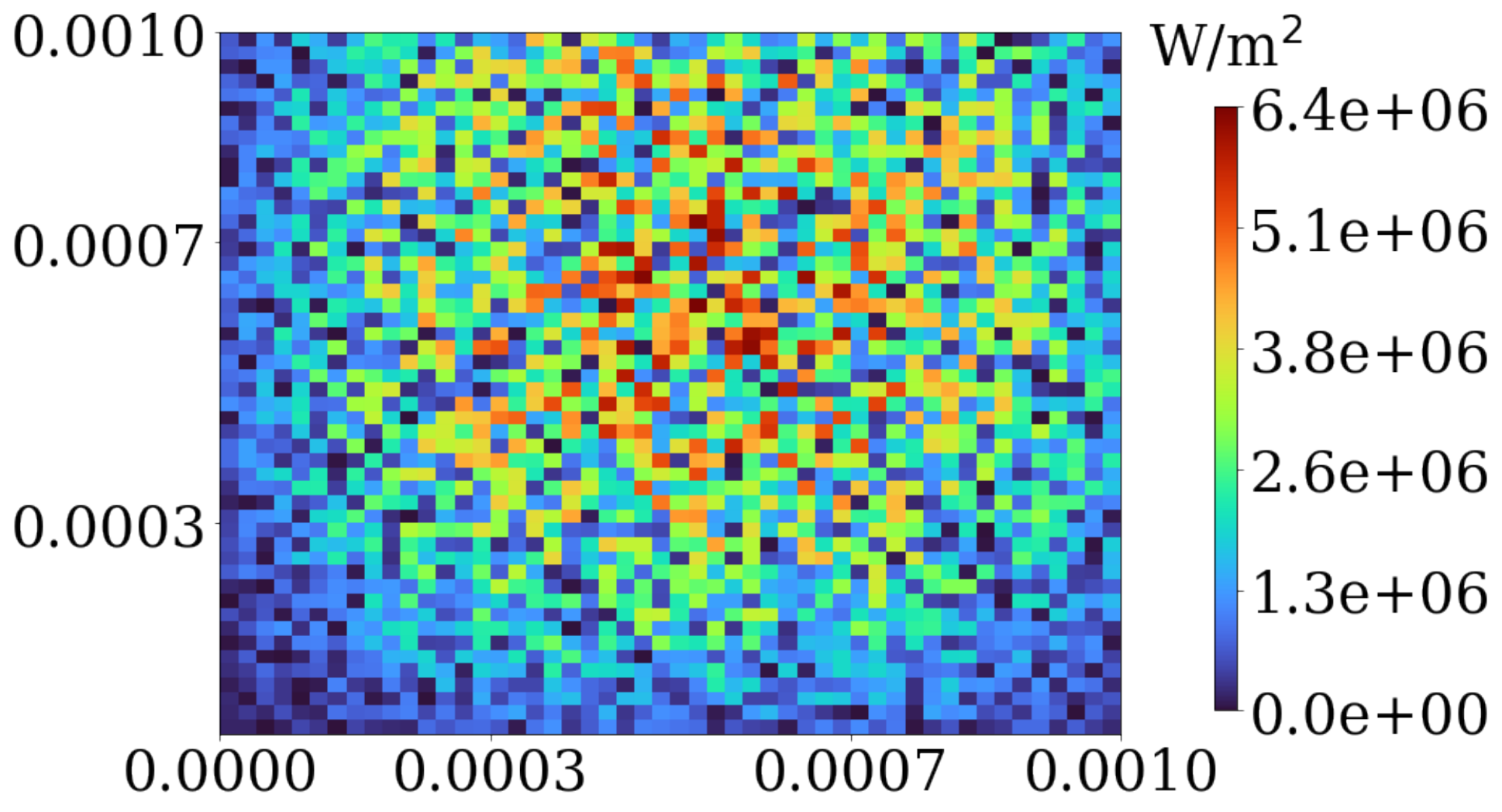}}}
    & {\raisebox{-.4\height}{\includegraphics[height=0.5in]{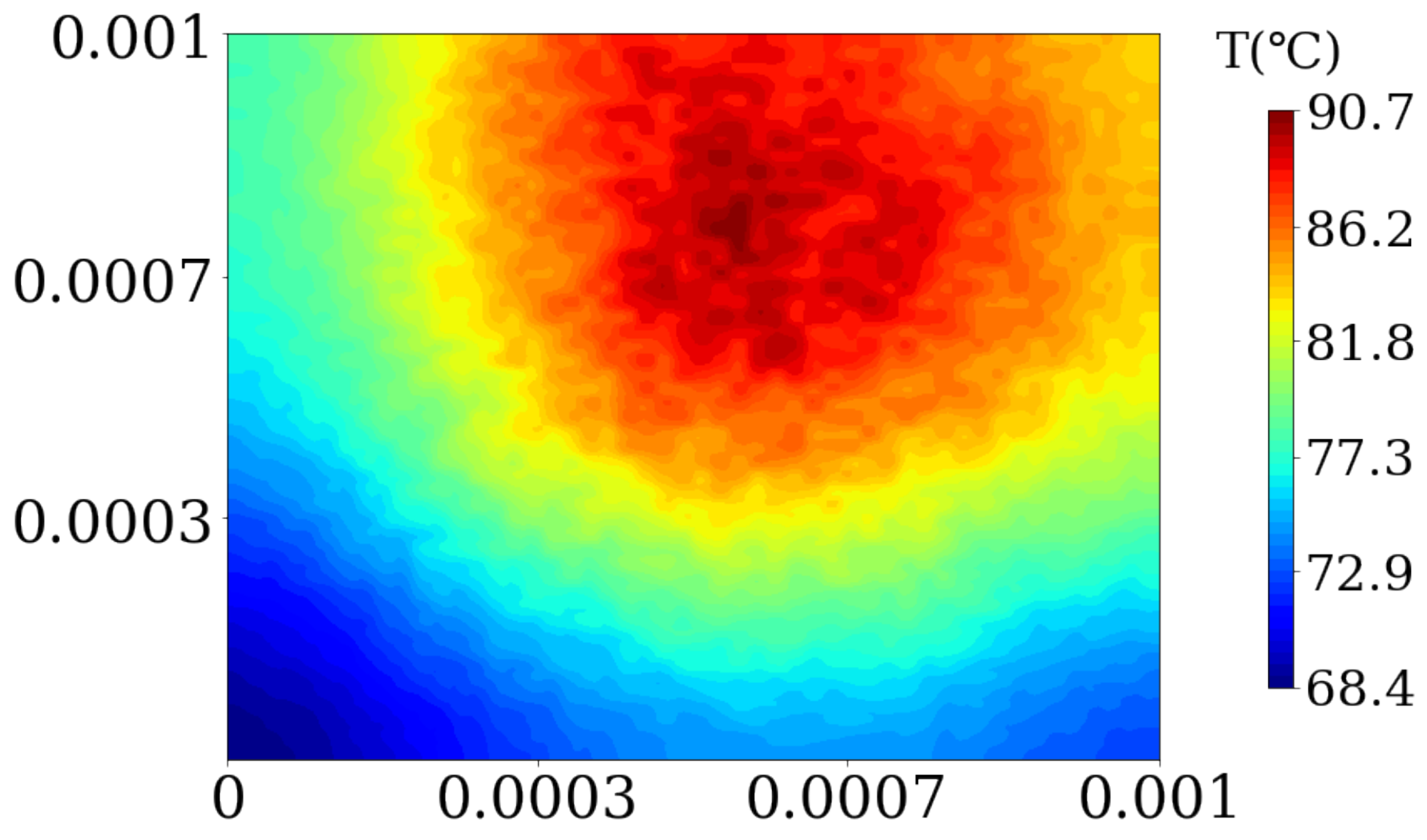}}}
    & {\raisebox{-.4\height}{\includegraphics[height=0.5in]{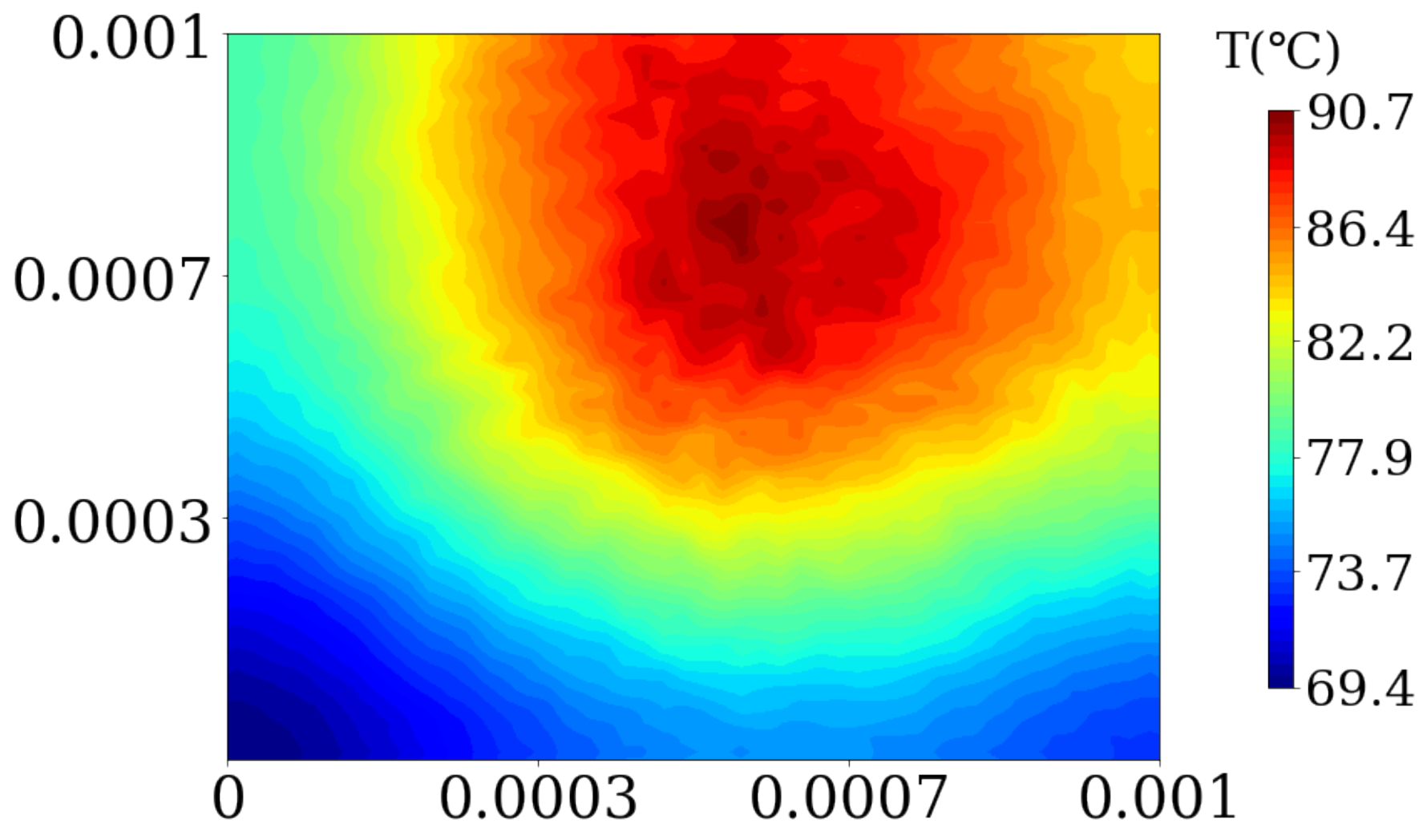}}} 
    & {\raisebox{-.4\height}{\includegraphics[height=0.5in]{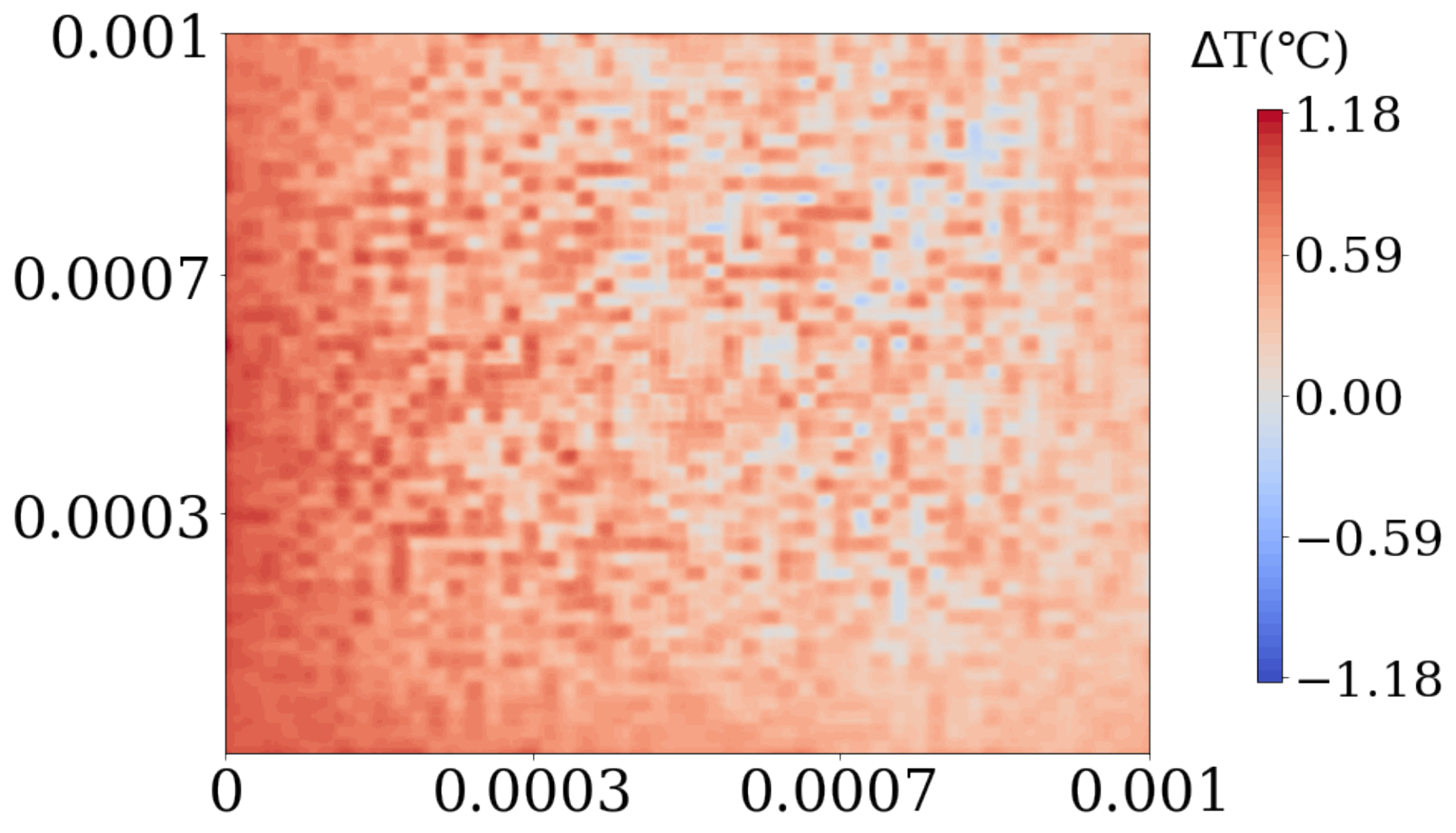}}}\\ [0.5cm] \midrule
    \textbf{TSV3D-IV}  
    & {\raisebox{-.4\height}{\includegraphics[height=0.5in]{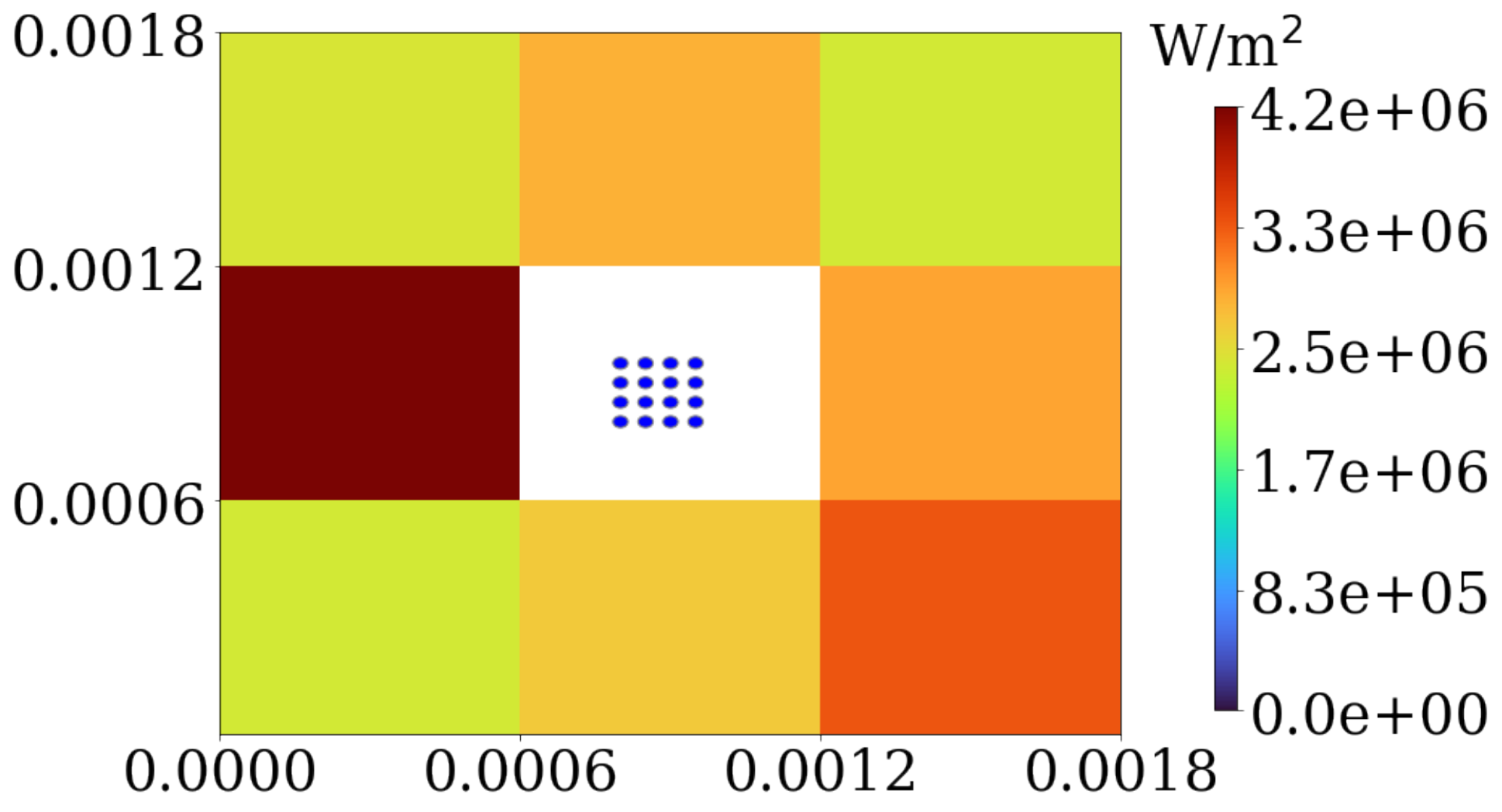}}}
    & {\raisebox{-.4\height}{\includegraphics[height=0.5in]{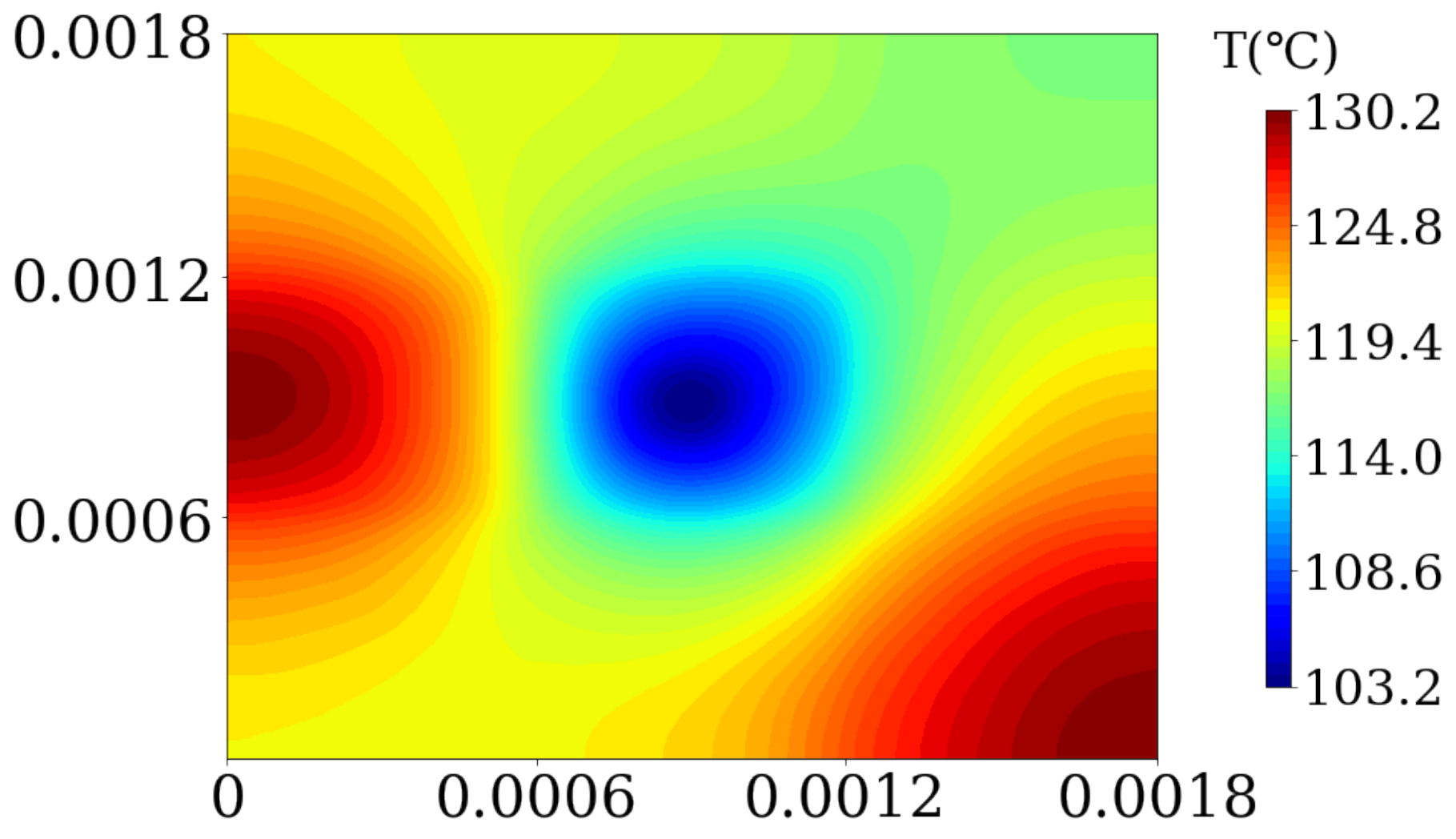}}}
    & {\raisebox{-.4\height}{\includegraphics[height=0.5in]{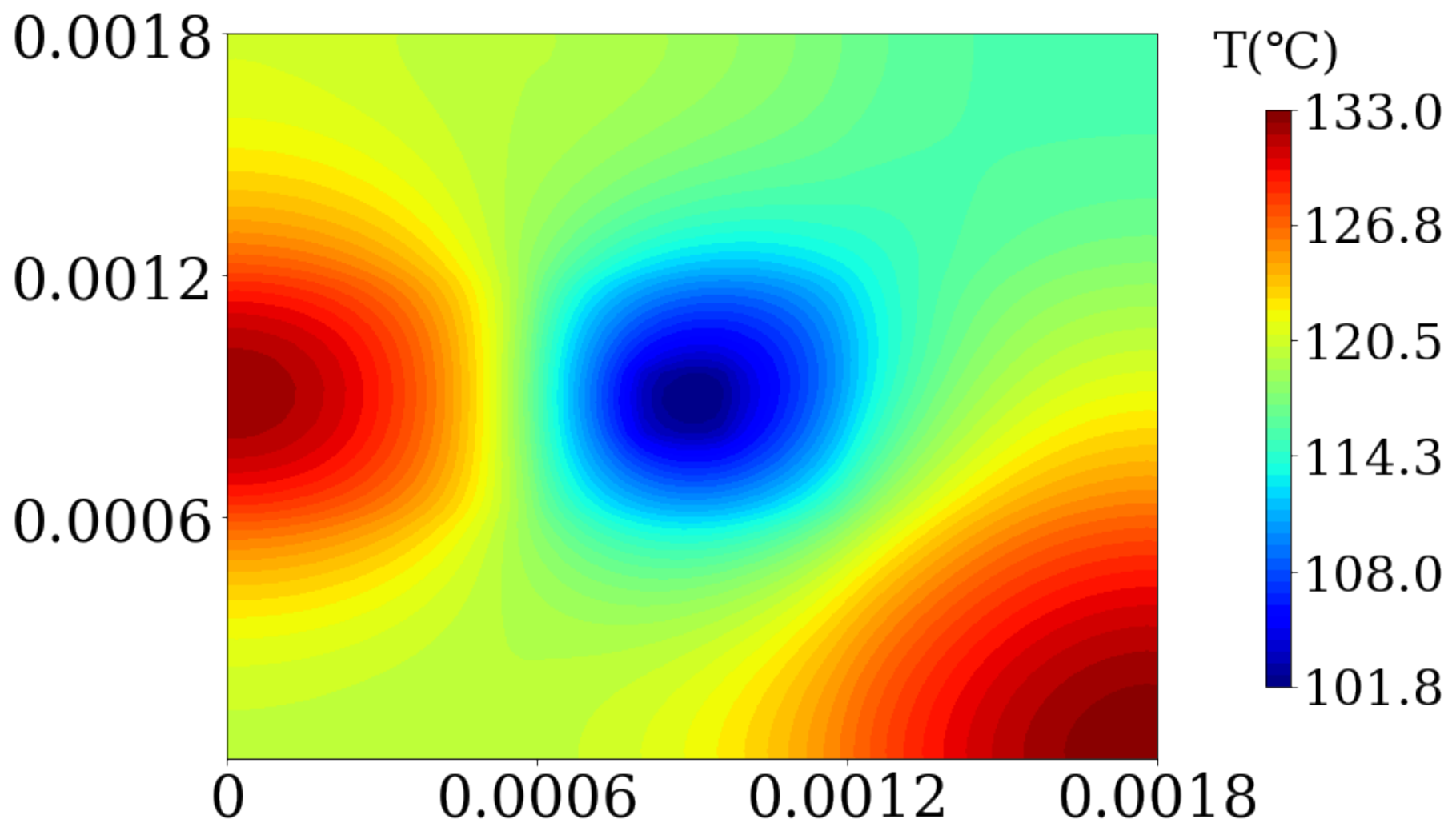}}} 
    & {\raisebox{-.4\height}{\includegraphics[height=0.5in]{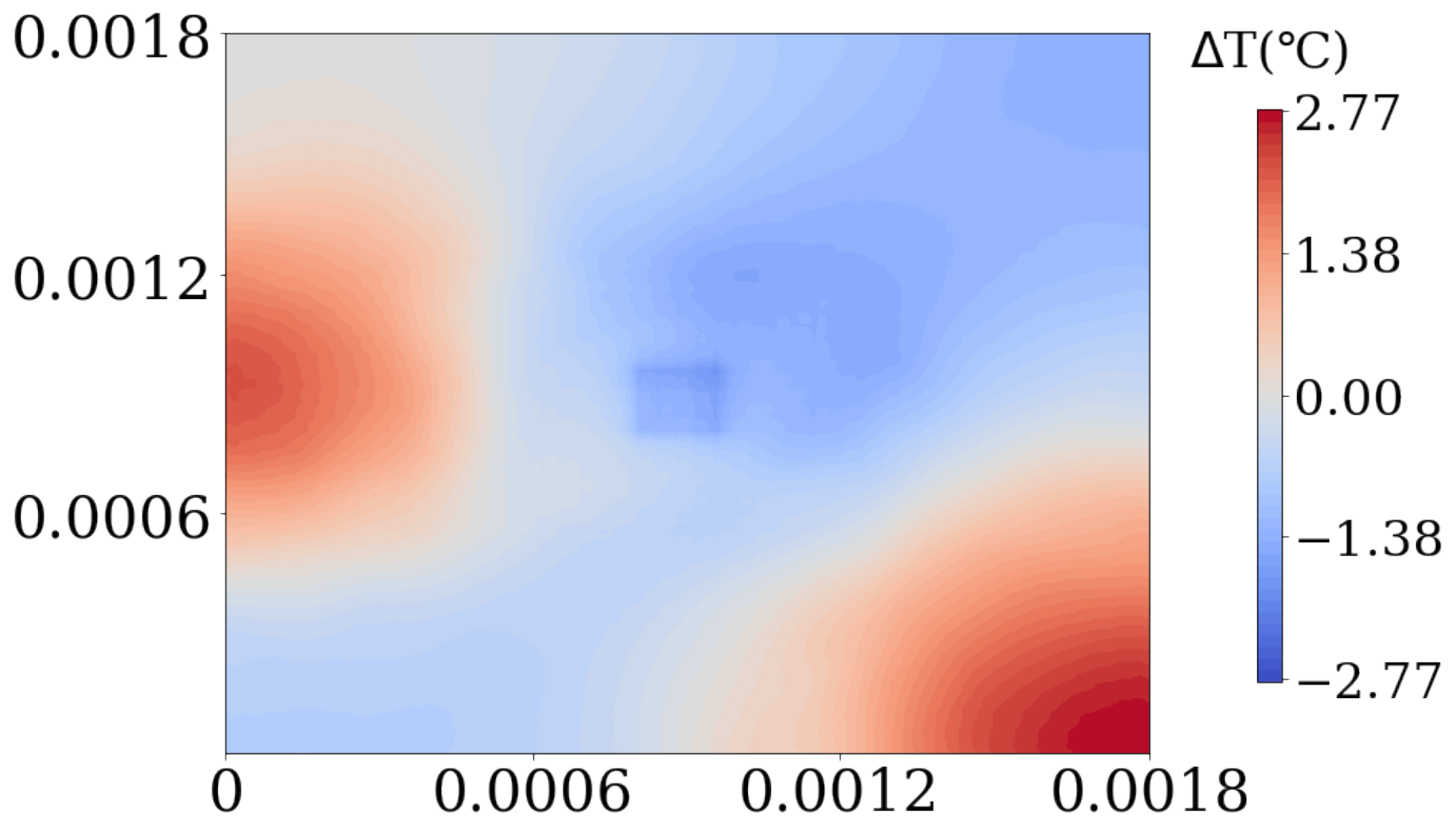}}}\\ [0.5cm]
    \bottomrule
\end{tabular}
}
\vspace{-0.2cm}
\label{table:atfig}
\end{table*}

In the following, we illustrate in Fig. \ref{Fig:runtime} the relation between the runtime in \textbf{Mono3D-I} and the number of cores employed, as well as the grid resolutions. The runtime first decreases and then saturates at a large number of cores (>16), as shown in Fig. \ref{Fig:runtime}(a). This comes from the fact that runtime for the local simulation decreases while the cost of creating, managing, and terminating the process pools increases when more cores are used.
Fig. \ref{Fig:runtime}(b) shows the relation between the runtime and global grid resolution for fixed local resolution. With the increase of global grid resolution, the global simulation time increases, the number of local subdomains increases, and the runtime for solving each sub-problem first decreases and then saturates. These factors collectively lead to the runtime exhibiting the trends in the figure. What's more, the error gets smaller with the increase in the number of global grids, indicating a trade-off between accuracy and efficiency when choosing the global grid resolution.

\vspace{-0.15cm}
\section{Conclusion}\label{sec:conclu}
As 3D-ICs are attracting more and more attention, related thermal simulation has met new challenges, including standard-cell level resolution, complex thermal effects (nonlinear leakage power and conductivity), and the heterogeneous material systems. However, current simulators require significant runtime for high-resolution simulation and dismiss these nonlinear effects. To this end, we propose ATSim3D, an accurate thermal simulator for steady-state simulation of nonlinear and heterogeneous 3D-IC systems. We utilize the global-local approach, combining a compact thermal model at the global level, and a finite volume method at the local level. We tackle the nonlinear effects by Kirchhoff transformation and iterative updating of leakage and conductivity distribution. By parallelization in the local level simulation, ATSim3D achieves an average speedup of $40\times$ compared to COMSOL, with the relative error <3\% and max error $<3^{\circ}C$ for the test structures of both 2D and 3D (Mono3D and TSV-based) ICs. A SOTA resolution of $4096\times4096$ is also demonstrated, along with a maximum acceleration of 40$\times$ compared to PACT.
We believe that our simulator holds the promise for enhancing future thermal-aware design in 3D-ICs.
\vspace{-0.15cm}

\section*{Acknowledge}

This work was supported in part by the National Science Foundations
of China (Grant No. 62125401, 62034007) and the 111 project (B18001).

\small
\bibliographystyle{IEEEtran}
\bibliography{ref/ATSim.bib}
\end{document}